\documentclass[traditabstract]{aa} 
\usepackage[varg]{txfonts}
\usepackage{graphicx,xcolor}

\usepackage{natbib}
\bibpunct{(}{)}{;}{a}{}{,}

\usepackage{txfonts}
\newcommand{\faper} {\hbox{$f_{{\rm aper}}$}}

\newcommand{\halpha}{H$\alpha$}

\newcommand{\msun}{M$_\odot$}
\newcommand{\mi}{$\mu$m}
\newcommand{\kms}{km~s$^{-1}$}
\newcommand{\htwo} {\hbox{H$_2$}}
\newcommand{\lk} {\hbox{$L_{\rm K}$}}
\newcommand{\lwone} {$L_{\rm W1}$ }

\newcommand{\lwfour} {$L_{\rm W4}$ }
\newcommand{\mhtwo} {\hbox{$M_{{\rm H}_2}$}}
\newcommand{\mmol} {\hbox{$M_{{\rm mol}}$}}
\newcommand{\mmolcenter} {\hbox{$M_{{\rm mol,0}}$}}

\newcommand{\mstar} {\hbox{$M_{{\rm *}}$}}

\newcommand{\spitzer}{{\it Spitzer}}

\newcommand{\wise}{WISE}

\newcommand{\figuredir}{figures/spectra}

\begin{document}

   \title{The role of molecular gas in galaxy transition in compact groups}


   \author{U. Lisenfeld
          \inst{1}
          \and
          K. Alatalo\inst{2}  
        \and
          C. Zucker\inst{3}  
          \and
        P. N. Appleton\inst{4}
          \and
        S. Gallagher\inst{5,6}
          \and
         P. Guillard\inst{7,8}
         \and
        K. Johnson\inst{9}
        }

   \institute{Departamento de F\'isica Te\'orica y del Cosmos, Universidad de Granada, Spain and Instituto Carlos I de F\'isica T\'eorica y Computacional, 
   Facultad de Ciencias, 18071 Granada, Spain\\             
    \email{ute@ugr.es}
         \and
            Observatories of the Carnegie Institution of Washington, 813 Santa Barbara Street, Pasadena, CA 91101, USA
             \and
             Harvard-Smithsonian Center for Astrophysics, Cambridge, MA 02138, USA
             \and
   NASA Herschel Science Center, IPAC, Caltech, Pasadena, CA 91125, USA
   \and
   Department of Physics and Astronomy, University of Western Ontario, London, ON N6A 3K7, Canada
   \and
   Centre for Planetary and Space Exploration, University of Western Ontario, London, ON N6A 3K7, Canada
   \and
   Institut d'Astrophysique de Paris, CNRS, UMR 7095, 98 bis Boulevard Arago, F-75014 Paris, France
   \and
Sorbonne Universit\'es, UPMC Universit\'e Paris 06, 4 place Jussieu, F-75005 Paris, France
\and
Department of Astronomy, University of Virginia, P.O. Box 400325, Charlottesville, VA 22904-4325, USA
             }

   \date{Received September 15, 1996; accepted March 16, 1997}

 
  \abstract
  {
  Compact groups (CGs) provide an environment in which interactions between galaxies and with the intra-group medium
  enable and accelerate galaxy transitions from actively star forming  to quiescent.  Galaxies in  transition  
   from active to quiescent can
be selected,   by their infrared (IR) colors, as canyon   or 
  infrared transition zone (IRTZ) galaxies.
We used a sample of CG  galaxies with IR data from the  {\it Wide Field Infrared Survey Explorer} (\wise)
 allowing us to calculate the stellar mass and star formation rate (SFR) for each galaxy. 
 Furthermore, we present new CO(1-0)  data for { 27 galaxies and
   collect data  from the literature}  to calculate the molecular gas mass
  for  a total sample of 130 galaxies.  This data set allows us 
 to study the difference in the molecular gas fraction (\mmol/\mstar) and star formation efficiency (SFE=SFR/\mmol) between
  active, quiescent, and transitioning (i.e., canyon and  IRTZ) galaxies.
    We find that transitioning   galaxies have a mean molecular gas fraction and a mean SFE  
    that are significantly  lower than those of actively star-forming 
    galaxies. The molecular gas fraction is higher than that of quiescent galaxies, whereas the SFE 
    is similar.
    These results indicate that the transition from actively star-forming to quiescent in CG galaxies
     goes along with a loss of molecular gas, possibly due to tidal forces exerted from the
    neighboring galaxies or a decrease in the gas density. 
    In addition, the remaining molecular gas loses its ability to form stars efficiently, possibly { owing to 
   turbulence perturbing the gas, }
as seen in other, well-studied examples such as Stephan's Quintet and HCG~57.  
Thus,  the amount and properties of molecular gas play a crucial  role in the environmentally driven transition of galaxies from actively star forming
to quiescent.
  }


   \maketitle
%

\section{Introduction}

Galaxies show a bimodal distribution in optical color space, most of which are either in a blue cloud
or a red sequence \cite[e.g.,][]{baldry04}. The dearth of galaxies in the green valley, in between the red and blue galaxies,
 has been
interpreted as a rapid transition between the two phases \citep{bell04,faber07}. 
%
The mechanisms responsible for this transition are not clear and there are most likely different causes 
as seen, for example, in the analysis of \citet{schawinski14}. These authors demonstrated
 that the green valley does not consist of a homogeneous
group of galaxies, but is composed of 
fast-evolving early-type galaxies and secularly evolving late-type galaxies.
This shows that 
optical colors alone are not enough to study galaxy transitions.
 
Infrared (IR) colors of galaxies can give complementary information.  Based on WISE colors,
\citet{alatalo14b} { found 
a prominent bifurcation between late-type and early-type galaxies in the  \wise\ color  range of
0.8 $\lesssim$ W2-W3 $\lesssim$  2.4, which they
labeled the infrared transition zone (IRTZ).  }
Interestingly, the IRTZ does not coincide with the optical green valley. Instead, galaxies
in the IRTZ tend to have mostly red optical  colors and 
are predominantly early-type galaxies. This indicates that galaxies seem to pass the optical green valley first, and later the
IRTZ, possibly by completely shedding their interstellar medium (ISM) \citep{alatalo14b}. 
A similar result was found by \citet{walker13} for galaxies in compact groups (CG).

In an  analysis of \spitzer\ Space Observatory IRAC colors, \citet{johnson07} found an IR gap, 
i.e., a seemingly underpopulated region, in Hickson compact groups (HCGs),
 separating actively star-forming  galaxies from  galaxies dominated by an evolved stellar population.
 The low density of galaxies in this gap suggests that the evolutionary phase corresponding to the gap is a very short transition. 
 The existence of an underpopulated region was confirmed in larger samples \citep{walker10} but 
 confined to a smaller region in mid-IR color space
  \citep{walker12}. 
  { With star formation rates (SFR) and stellar masses measured from UV and near- and mid-IR photometry, \citet{tzanavaris10} identified
  specific star formation rate (sSFR) as the physical driver for location in mid-IR color space.}
 
 \citet{zucker16} expanded these studies, based on a limited amount of \spitzer\ data,
 using WISE data. They measured the fluxes in all four WISE bands for a sample of 652 galaxies in 163 CGs and classified these galaxies, using colors,  
 into mid-IR active (meaning actively star-forming), mid-IR quiescent, and mid-IR canyon galaxies\footnote{
{ In the most parts of this paper, and following \citet{zucker16}, we call these groups simply active, quiescent and canyon
galaxies.}}.
%
 Their method consisted in  plotting the galaxies in a unique region of WISE color-color space log(W3-W2) vs. log(W4-W1) to separate galaxies dominated by polyaromatic hydrocarbon (PAH) emission from those dominated by stellar light. Zucker et al. then applied 2D iso-density contours to identify canyon galaxies occupying the underdense region between the active and quiescent populations.
 These authors showed that this classification is compatible with the previous classification based on \spitzer\ data, 
 allowing the enlargement of  the sample by a factor of three owing to the complete sky coverage of the WISE observations.
 Galaxies classified as belonging to the canyon between active and quiescent also exhibit a sSFR between active and quiescent galaxies, implying that this
population represents a transition phase between the two in which the star formation (SF) is quenched.
 
Additional indications that something unusual is going on in these galaxies was found by \citet{cluver13}  who 
observed  a sample of Hickson compact groups (HCGs) with intermediate HI deficiencies with the \spitzer\ IRS
spectrograph. These authors identified a number of galaxies with enhanced warm H$_2$ emission that is well above the level expected from SF and indicative of shock excitation, classifying these galaxies as so-called molecular hydrogen emission galaxies \citep[MOHEGs; ][]{ogle10}. Interestingly, most of these galaxies have IRAC colors in the range of the previously found gap \citep{johnson07,walker10}. Furthermore, based on their extinction-corrected optical colors, these galaxies fall predominantly in the optical green valley between blue, star-forming galaxies 
and red-sequence objects. \citet{cluver13} conclude that the increased warm H$_2$ emission is most likely due to shock excitation.

The cause of the transition between active  and quiescent  is still unclear. The transition
goes in parallel with a decrease in the SFR, yet it is unclear what exactly shuts the 
SF down. In order to shed further light on this question, we need to quantify the SFR and also measure and characterize the molecular gas content that is available. Previous surveys of the
molecular gas \citep{leon98,verdes-montenegro98,martinez12}  found no deficiency of the molecular gas compared to field galaxies. 
\citet{lisenfeld14} measured the molecular gas content in a small sample of {31 HCG galaxies; they classified most of these with available data for the 
warm H$_2$ emission} and classified 14 as MOHEGs. Even though there were some objects with a very low SFE,
there was no clear relation to the intensity of the warm H$_2$ emission; instead, on average the SFE in MOHEG galaxies  was the same as in non-MOHEG  galaxies.
On the other hand, the  \mhtwo/\lk\ ratio was lower for MOHEG galaxies, indicating a decrease in the molecular gas content.
Some of the galaxies (mostly MOHEGs) showed very broad lines ($\sim$ 1000 \kms) and irregular line shapes that are indicative of
a perturbation of the molecular gas. The presence of kinematically disturbed molecular gas was confirmed with 
 CARMA observations  in HCG 57 \citep{alatalo14a} and later in a sample of 
14 HCGs \citep{alatalo15}, in which  a considerable SF suppression  was found in some objects, 
especially in those galaxies that were
in a more advanced stage of transition based on  their optical and IR colors.

The goal of the present paper is to clarify the role that molecular gas plays  in the
decrease of SF that goes along with the transition from active to quiescent. 
We approach this question based on a sample of CG galaxies. 
Because of their low velocity dispersions and close proximity to other group members, 
these galaxies are strongly affected by interactions with their companions and the intra-group medium. 
The effects are clearly visible as evidenced by a large percentage of the atomic hydrogen (HI) that is outside the galaxies
i{ n some groups}  \citep[e.g.,][]{verdes-montenegro01} and by the  morphological dominance of S0 galaxies.
We use the sample of 
\citet{zucker16} to study the relation between molecular gas, SF, and stellar mass as a function of
IR color. 
The sample and the CO data presented in this paper allow us  to answer the fundamental question of whether there are differences in the mean molecular gas fraction  and  star formation efficiency (SFE;  defined as the SFR  per molecular gas mass) 
between active, quiescent, and transitioning galaxies. { The classification into these groups is
based on their mid-IR colors}. We consider two different groups of transitioning galaxies:
canyon galaxies as defined by \citet{zucker16} and  IRTZ galaxies as defined by \citet{alatalo14b}. 


\section{Sample and data}
\label{sec:sample_data}

Our sample is based on the catalog of \citet{zucker16}, which presents WISE data for 652 galaxies in 163 compact groups, of which 428 galaxies have reliable photometry (S/N $>$ 2 in all bands). 
The groups are composed of 93 HCGs, identified in a systematic visual search by \citet{hickson82} and later cleaned up with the availability of velocity data  \citep{hickson92}. 
In addition, the sample includes 70 groups from the Redshift Survey Compact Group catalog \citep[RSCG;][]{barton96} whose galaxies were chosen from a
 complete, magnitude-limited redshift survey via a friends-of-friends algorithm. \citet{zucker16}  excluded discordant HCG groups and galaxies, checked the two samples 
 for overlap, and excluded RSCGs that were embedded in clusters. In addition to this and following  \citet{barton96}, {we excluded 
 very close-by systems (cz  $\le 2300$ \kms), whose environment cannot be analyzed reliably due to their large angular sizes.} 
  We also implemented a stricter signal-to-noise cut than in \citet{zucker16} and excluded any galaxies with S/N $<$ 2.5 in lieu of the S/N $<$ 2 cut they used. In all cases, the W4 band was the limiting factor for this criterion; i.e., if a galaxy was detected in W4, it was detected in the other bands as well. 
 Applying both the redshift and signal-to-noise cuts, we narrow our potential sample down to 294 of 428 galaxies used in the \citet{zucker16} analysis. 

We searched the literature for all existing CO data for this 
\citet{zucker16} subsample (294 galaxies) and obtained CO measurements for { 102}  HCG galaxies 
\citep{verdes-montenegro98,leon98,martinez12,lisenfeld14} and for two RSCG galaxies  \citep[NGC~232 and NGC~2831]{mirabel90,wiklind95}\footnote{
 { Three galaxies that were all classified as active by
\citet{zucker16}, with available CO and WISE data, 
were not included in the present study. This is because of possible contamination of the mid-IR emission by an AGN  \citep[HCG~56b
and HCG~96a;][]{cluver13}
and an unusually high SFE   in the merging
object HCG~31ac that is possibly due to an intergalactic starburst, making it
atypical for our study \citep[see][for a discussion]{lisenfeld14}}.}.
The observations were carried out with the Institut de Radioastronomie Millimetrique (IRAM) 30 m telescope, Five College Radio Telescope, Swedish-ESO Submillimetre Telescope (SEST), and Kitt Peak Radio Telescope with single 
pointings at the central position for most cases. { To supplement the CO data for these 104 galaxies  from the literature, as part of this study we observed the redshifted CO(1-0) 
line for an additional 27 galaxies. The details of these observations are described  in section~\ref{sec:co_obs} below.
{ One of the objects, HCG~54a, was later excluded from the statistical analysis because of its low
recession velocity, but we present the CO data here for completeness.}
 In total, our final sample includes CO data for 130 galaxies (104 from the literature and 26 obtained herein, of which 89 are detections), which constitutes 
 about one-third of the reliable photometric 
 sample of \citet{zucker16}. 
 
Our selection of galaxies from the sample of \citet{zucker16} was
driven by the availability of CO data. Of the 294 galaxies from \citet{zucker16} 
with reliable mid-IR photometry (S/N > 2.5 in all four
WISE bands), only 130  had available CO data.
While we are not affected by any
obvious biases because each CO study from the literature used
different selection criteria, we nevertheless need to check that our
results using this 130 galaxy subsample are robust. We carried out
this test by creating an additional subsample that only contained galaxies in those groups with complete CO coverage, yielding 89 of 130
galaxies. Explicitly, we went through the \citet{zucker16} sample
group by group and determined whether all galaxies in that group with
reliable photometry (S/N > 2.5) also had CO data. If they did, all
reliable photometry galaxies in that group were included in the bias
test. Otherwise all galaxies were excluded. This allowed the compact
group environment to remain intact with respect to the \citet{zucker16}
sample, by avoiding the inclusion of only some galaxies in a
group that might have been preferentially selected for CO follow-up
observations. In this way, these 89 of 130 galaxies were selected
based on the W4 flux rather than CO. We carried out the entire
analysis for this smaller 89 galaxy subsample and obtained entirely
consistent results, indicating that the availability of CO data did
not affect our conclusions. The details are presented in Appendix A.

}

\subsection{CO observations}
\label{sec:co_obs}

We observed an additional { 27} galaxies in CGs between  January and April 2017 with the 
IRAM 30 m telescope on Pico Veleta. 
We selected the sources, based on their WISE colors, as preferentially canyon or IRTZ objects.
We observed the redshifted $^{12}$CO(1-0)  line 
with the  dual polarization receiver EMIR in combination  with 
the autocorrelator FTS at  a   frequency resolution of 0.195~MHz  (providing a velocity resolution of $\sim$ 0.5~\kms\  at CO(1--0))
and with the autocorrelator  WILMA with a frequency resolution of  2~MHz (providing a velocity resolution of $\sim$ 5~\kms\ at CO(1--0)).
The observations were carried out in wobbler switching mode with a
wobbler throw between 40\arcsec\ and 140\arcsec\ in azimuthal direction. The wobbler throw was chosen individually in order to 
ensure that the off position was away from neighboring galaxies. 

The  broad bandwidth of the receiver (16~GHz) and backends (8~GHz for the FTS and 4~GHz for WILMA) 
{ enabled grouping of observations}
of galaxies with similar redshifts. The central sky  frequencies, taking into account
the redshift of the objects, ranged between 108 and 112~GHz. 
Each object was observed until it was detected with a S/N  ratio of   $\sim$ 5 or
until the root-mean-square (rms) noise was below $\sim$ 3 mK (T$_{\rm A}^*$)  for a velocity resolution of 20  km s$^{-1}$.
The integration times per object ranged from 30 to 130 minutes. 
Pointing was
monitored on nearby quasars  every 60 -- 90  minutes.
During the observation period, the weather conditions were 
generally good with a pointing accuracy better than 3\arcsec.
The mean system temperature for the observations
was 160~K 
on the $T_{\rm A}^*$ scale.
At 115~GHz, the  IRAM forward
efficiency, $F_{\rm eff}$, is 0.95  and the 
beam efficiency, $B_{\rm eff}$, is 0.77.
The
half-power beam size is   22.3$^{\prime\prime}$ (for 110 GHz) . 
All CO spectra and luminosities are
presented on the main beam temperature scale ($T_{\rm mb}$), which is
defined as $T_{\rm mb} = (F_{\rm eff}/B_{\rm eff})\times T_{\rm A}^*$.

The data were reduced in the standard way via the CLASS software
in the GILDAS package\footnote{http://www.iram.fr/IRAMFR/GILDAS}.
We first discarded poor scans and then subtracted a constant or {  linear} baseline.
We then averaged the spectra and 
smoothed them  to resolutions between 20 and 40 \kms\ to increase
the signal-to-noise (S/N) ratio.

We present the detected spectra  in the appendix (Fig. A1).
For each spectrum, we determined visually the zero-level line widths, if detected. 
The velocity integrated spectra were calculated by summing the individual channels in between
these limits. 
 For non-detections we set an upper limit as

\begin{equation}
I_{\rm CO} < 3 \times rms \times \sqrt{\delta \rm{V} \ \Delta V},
\end{equation}

\noindent where $\delta \rm{V}$ is the channel width, $\Delta$V the zero-level line width, and $rms$ the root mean square noise.
For the non-detections, we assumed a linewidth of $\Delta$V = 300  \kms.
We treated tentative detections, with a S/N ratio between 4-5, as upper limits in the statistical analysis.
The results of our CO(1-0)  observations are listed in Table~\ref{tab:ico}.

\begin{table}
\caption{Integrated CO intensities}
\label{tab:ico}
\begin{tabular}{lccc}
\hline
Galaxy & rms\tablefootmark{a} & $ I_{\rm CO(1-0)}$  & $\triangle V_{\rm CO(1-0)}$\tablefootmark{b}  \\
  &  [mK]  & [K km s$^{-1}$] & [\kms] \\
\hline
    HCG01b &    1.58 &    1.41 $\pm$ 0.15 &    410 \\
    HCG05a &    1.51 &    5.73 $\pm$ 0.12 &    275 \\
    HCG19c &    3.24 &   $<$ 0.77 &    300\tablefootmark{c} \\
    HCG28a &    1.97 &    2.16 $\pm$ 0.22 &    570 \\
    HCG32d &    1.27 &    0.83 $\pm$ 0.14 &    590 \\
    HCG39a &    1.16 &    0.97 $\pm$ 0.15 &    800 \\
    HCG45a &    1.08 &    1.98 $\pm$ 0.17 &   1120 \\
    HCG45b &    1.08 &    0.62 $\pm$ 0.14* &    800 \\
    HCG46b &    1.35 &    0.91 $\pm$ 0.12 &    370 \\
    HCG47c &    1.89 &    2.47 $\pm$ 0.18 &    434 \\
    HCG51b &    2.90 &    3.58 $\pm$ 0.27 &    415 \\
    HCG52a &    1.32 &    0.79 $\pm$ 0.14 &    536 \\
    HCG54a\tablefootmark{d} &    3.37 &   $<$ 0.80 &    300 \\
    HCG57b &    1.87 &    1.37 $\pm$ 0.22 &    634 \\
    HCG57e &    1.85 &    1.22 $\pm$ 0.19 &    489 \\
    HCG60b &    0.93 &    1.09 $\pm$ 0.12 &    715 \\
    HCG64a &    2.44 &    1.71 $\pm$ 0.23 &    423 \\
    HCG70a &    2.12 &   $<$ 0.51 &    300\tablefootmark{c} \\
    HCG70b &    2.45 &    2.69 $\pm$ 0.27 &    564 \\
    HCG70c &    1.55 &    1.04 $\pm$ 0.14 &    375 \\
    HCG71a &    2.36 &    6.71 $\pm$ 0.24 &    470 \\
    HCG75c &    1.28 &    0.54 $\pm$ 0.12* &    389 \\
   NGC0070 &    3.53 &   $<$ 0.85 &    300\tablefootmark{c} \\
  NGC4410NED01 &    3.09 &    5.11 $\pm$ 0.35 &    604 \\
   NGC4614 &    1.94 &    1.56 $\pm$ 0.15 &    295 \\
   NGC6961 &    2.50 &   $<$ 0.60 &    300 \\
 MCG+02-36-018 &    1.75 &    0.73 $\pm$ 0.11 &    200 \\
\hline 
\end{tabular}
\tablefoot{
\tablefoottext{a}{Root-mean-square noise level at a velocity resolution of $\sim$ 21 \kms.}
\tablefoottext{b}{Zero-level line width. The uncertainty is roughly given by the velocity resolution.}
\tablefoottext{c} {Assumed linewidth for the non-detections.}
\tablefoottext{d} {{ HCG~54a was excluded from the subsequent statistical analysis because of its low
recession velocity.}}
\tablefoottext{*}{Tentative detection (S/N between 4 and 5). }
}
\end{table}

\subsection{Molecular gas mass}
\label{sec:molecar_gas_mass}

We calculated the molecular gas mass within the pointing, \mmolcenter,
from the CO(1-0) emission with the following equation:
\begin{equation}
M_{\rm mol,0} [M_\odot]=102 \times D^{2}I_{\rm CO(1-0)}\Omega
\label{eq:mh2}
.\end{equation}


Here,  $\Omega$ is the area covered by the observations in arcsec$^{2}$ (i.e., $\Omega$ = 1.13 $\theta^{2}$ 
for a single pointing with a Gaussian beam where $\theta$ is the HPBW), $D$ is the distance in Mpc, and $I_{\rm CO(1-0)}$ is the velocity integrated line intensity in
K \kms . This equation assumes a CO-to-H$_2$ conversion factor 
 X=$N_{\rm H_{2}}/I_{\rm CO}$ = $2\times 10^{20}\rm cm^{-2}$ (K km s$^{-1})^{-1}$ (Bolatto et al. 2013)
and includes a factor of 1.36 to account for helium and other heavy metals.

The CO data in our sample were obtained with different telescopes and  different HPBWs.
Furthermore, only the central positions were observed. Therefore, we need to apply a correction for emission
outside the beam to derive the total molecular gas mass, which is necessary to be able to compare the
data from different telescopes.
We carried out this aperture correction in the same way as described in \citet{lisenfeld11}, 
assuming an exponential distribution of the CO emission,

\begin{equation}
I_{\rm CO}(r) = I_{0}\cdot \exp(r/r_{\rm e}) .
\label{eq:Ico_r}
\end{equation}

We adopted a scale length of $r_{\rm e}$ = 0.2$\times$$r_{\rm 25}$, where $r_{\rm 25}$  is the major optical isophotal radius at 25 mag arcsec$^{-2}$, 
following \citet{lisenfeld11}, who derived this scale length from different  studies 
 \citep{2001PASJ...53..757N, 2001ApJ...561..218R, 2008AJ....136.2782L}
 and from their own CO data.  These studies are based on spiral galaxies, however, the analysis of the spatial extent of CO in early-type galaxies in the 
 ALTAS$^{\rm 3D}$ survey showed that the relative CO extent, normalized to $r_{\rm 25}$,  of early-type and spiral galaxies is  the same
 \citep{2013MNRAS.429..534D}. 
We used this distribution to calculate the expected CO emission from the entire disk, taking the galaxy inclination into account
 \citep[see][for more details]{lisenfeld11}. 
We obtained both the isophotal radius (corrected for  galactic extinction and inclination) and the inclination from the 
Hyperleda\footnote{http://leda.univ-lyon1.fr/} 
 database \citep{markov14}.

The resulting aperture correction factor, \faper, defined as the
ratio between the molecular gas mass observed in the central
pointing, \mmolcenter\ and the molecular gas mass extrapolated to the
entire disk, \mmol, lies between 1.0 and 5.3 with a mean value
of 1.7. The values for the molecular gas mass in the central
pointing and the extrapolated molecular gas mass are listed in Table~\ref{tab:mh2_sfr_mstar}.

The empirical aperture correction could have potentially introduced biases in our analysis.
{We 
carried out two bias tests, 
described in detail in Appendix~\ref{sect:aperture_corr}, even though} the small mean value of \faper\  (1.7),  which is considerably less than
the  differences in  SFE and \mmol/\mstar\ found between the galaxy groups  (factors $>$ 3), 
makes a strong bias unlikely.
Firstly, we compared the
distribution of \faper\ for the different galaxy groups studied in this analysis (active,
quiescent, canyon, and IRTZ galaxies) and found that 
there is no significant difference.
Secondly, we carried out our
entire analysis for a subsample of galaxies with a small  aperture correction
(\faper $< 1.6$). We found that the results are entirely consistent with the analysis of the
full sample. Thus, we conclude that the application of the aperture correction did
not bias our results.

\begin{table*}
\caption{Molecular gas mass, SFR, and stellar mass}
\label{tab:mh2_sfr_mstar}
\begin{tabular}{lcccccc}
\hline
Galaxy & Distance\tablefootmark{a}  & log(\mmolcenter)\tablefootmark{b} &log(\mmol)\tablefootmark{b} & Ref. \tablefootmark{d}& SFR\tablefootmark{e}  & log(\mstar)\tablefootmark{f}  \\
   &  [Mpc]  & [\msun] &  [\msun] &  &  [\msun\ yr$^{-1}$] &  [\msun]  \\
\hline
   HCG01b&   146.6&   9.23&   9.35&   1&    0.63&  10.61\\
     HCG02b&    62.1&   8.92&   9.19&   4&    6.08&   9.65\\
     HCG04a&   114.9&  10.09&  10.49&   4&   23.45&  10.49\\
     HCG05a&   174.0&   9.99&  10.24&   1&    1.60&  10.99\\
     ..... & ..... &  ..... & .....& ....& ....& ....\\
\hline 
\end{tabular}
\tablefoot{
\tablefoottext{b}{Distance derived from redshift with $H_0 = 70$~\kms\ Mpc$^{-1}$.}
\tablefoottext{b}{Cold molecular gas within the central pointing, calculated from eq.~\ref{eq:mh2}.}
\tablefoottext{c}{Cold molecular gas mass, extrapolated from the central pointing to the entire disk, calculated as
described in Sect.~\ref{sec:molecar_gas_mass}.}
\tablefoottext{d}{References for the CO(1-0) measurements:
(1) this work; (2) \citet{lisenfeld14}; (3) \citet{martinez12}; (4) \citet{leon98}; (5) \citet{verdes-montenegro98};
(6) \citet{wiklind95}; and  (7)  \citet{mirabel90}.}
\tablefoottext{e} {Star formation rate, calculated from the WISE W4 band with eq.~\ref{eq:sfr}.}
\tablefoottext{f} {Stellar mass, calculated from the WISE W1 and W2 bands with eq.~\ref{eq:mstar}.}

The full table is available in electron form at the CDS.
}
\end{table*}

\subsection{Star formation rate}
   
The mid-IR emission is a good tracer for the SFR. 
We used the WISE data in band 4 (W4, $\lambda = 22$ \mi) from \citet{zucker16} to calculate the SFR. We adopted the prescription of  
\citet{lee13} who calibrated this band as a SFR tracer  for a sample of $\sim$ 100\,000 
galaxies retrieved from the SDSS, based on a comparison with  the extinction corrected \halpha\ luminosity.
These authors obtained (after adapting to a Kroupa IMF by multiplying by 1.59) a relation

\begin{equation}
SFR =  10^{-9} L_{\rm W4} [M_\odot yr^{-1}],
\label{eq:sfr}
\end{equation}
where \lwfour = $\nu f_{\rm W4} 4\pi D^2$ is the luminosity in the W4 band { (in units  of bolometric solar luminosity)}, 
where  f$_{\rm W4}$ is the flux in this band.
This relation is similar to that found by \citet{jarrett13}.
The values for the SFR are listed in Table~\ref{tab:mh2_sfr_mstar}.

\subsection{Stellar mass}

The near-IR wavelength range  is dominated by emission from low-mass stars and is therefore a very good measure
of the total stellar mass.
We calculated the stellar mass from the luminosity  in the WISE W1 and W2 bands (at 3.4~\mi\ and 4.6~\mi), adopting eq. (9) from   
\citet{jarrett13}, i.e.,

\begin{equation}
\log\left(\frac{\mstar}{L_{\rm W1} }\right) = -0.246 - 2.100(W1-W2).
\label{eq:mstar}
 \end{equation}
where W1 and W2 are given in magnitudes and \lwone\ is the total in-band luminosity, derived by multiplying the spectral luminosity
\lwone= $\nu f_{\rm W1} 4\pi D^2$  by a factor of 22.883.  As explained in \citet{jarrett13}, this factor accounts for the difference
between the total solar luminosity and the in-band values as measured by WISE.
We converted from the flux densities listed in \citet{zucker16} to magnitude using the relationship 
$m_{VEGA} = - 2.5 \log\left( \frac{f_\nu}{f_{\nu,0}}\right)$ with $f_{\nu,0}$ given by 306,682, 170.663, 29.0448, and 8.2839 for bands $W1, 
W2, W3,$ and $W4$, respectively,
as outlined in Section IV.4 of the \wise\ All-sky Explanatory Supplement. 
The values of the stellar mass are listed in Table~\ref{tab:mh2_sfr_mstar}.



\section{Results}

The available data allow us to study the molecular gas fraction (\mmol/\mstar) and the star formation efficiency
(SFE = SFR/\mmol) of the galaxies as a function of their activity class. For this, we distinguish galaxies based on their
WISE classification from \citet[][mid-IR active, mid-IR quiescent, and mid-IR canyon galaxies]{zucker16} 
 and search for differences between the three groups.

We also investigate the properties of galaxies lying in the IRTZ, defined by 
 \citet{alatalo14b}  as galaxies with \wise\ colors in the range
0.8 $<$ W2(4.6\mi) -- W3(12 \mi) $<$  2.4, corresponding to the  transition region  between active and
quiescent galaxies.  The data by \citet{zucker16} showed  a systematic offset 
of 0.3-0.4~mag  in W2-W3 compared to the Alatalo sample \citep[see Fig.~7 of][]{zucker16}.
Zucker et al. argued that this offset is due to the different apertures  used in the photometry applied to each sample.
 Alatalo et al. 2014b used wgmag values from the ALLWISE source catalog, which 
is based on elliptical apertures scaled from the 2MASS  extended source catalog apertures, whereas Zucker et al. used
$1\sigma-3\sigma$ contours 
derived from the averaged, $\lambda^{-1}$  weighted reference image of the first three WISE bands.
The same offset was found by  \citet{cluver14} 
 between wgmag values and the use of isophotal contours.
Taking this offset into account, the IRTZ defined by \citet{alatalo14b} shifts to 1.2 $<$ W2(4.6\mi) -- W1(3.4 \mi) $<$  2.8.
We furthermore restrict the IRTZ to a more central zone in the W2-W3 distribution  with a minimum in galaxy density
between active and quiescent galaxies \citep[see Fig.~7 in][]{zucker16}.
Based on this figure, we define galaxies  in the IRTZ as those with 
1.4 $<$ W2-- W3 $<$  2.6.


\subsection{Molecular gas mass and  stellar mass}

\begin{table}
\caption{Mean  {  and median   log(\mmol/\mstar)} for different samples}
\begin{tabular}{lccc}
\hline
Class & mean & median & n/n$_{\rm up}$\tablefootmark{a} \\
\hline
{Full sample} & & \\
Active    & -0.67 $\pm$ 0.06  & -0.74  & 68/17 \\
Canyon  &  -1.13$\pm$ 0.07 & -1.20  & 24/8\\
IRTZ & -1.44$\pm$ 0.10 & -1.47  & 38/10 \\
quiescent  &  -1.91 $\pm$ 0.09 & -1.88 & 38/16  \\
\hline
{Late-types galaxies ($T > 0$)} & & \\
Active    & -0.67 $\pm$ 0.06  & -0.73 & 52/12 \\
Canyon  &  -1.04 $\pm$ 0.06 & -1.10 & 16/5 \\
IRTZ & -1.21$\pm$ 0.11 & -1.31 &  20/5 \\
quiescent  &  -1.50 $\pm$ 0.07 & -1.52 & 13/3 \\
\hline 
\end{tabular}
\tablefoot{
\tablefoottext{a} {Total number of galaxies (n) and number of upper limits (n$_{\rm up}$)}.
}
\label{tab:mean_mmol_over_mstar}
\end{table}

   \begin{figure}
   \centering
 \includegraphics[width=8cm,angle=270]{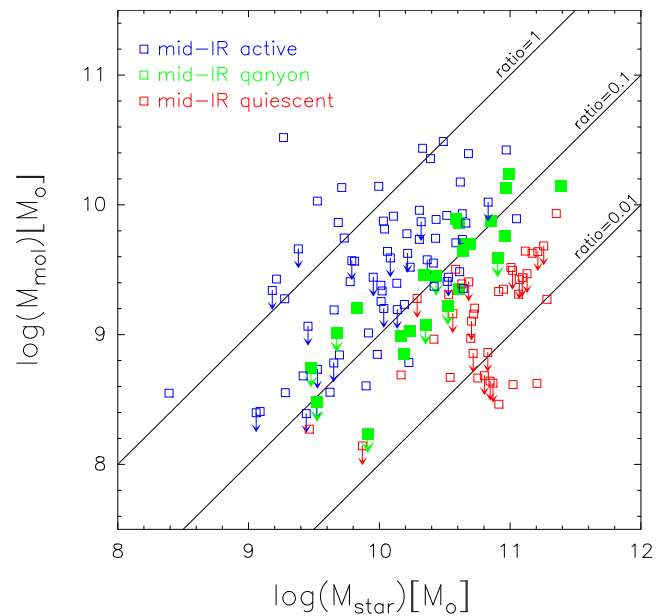}
      \caption{Stellar mass, derived with eq.~\ref{eq:mstar}, as a function of molecular gas mass.
      The lines show ratios of \mmol/\mstar = 100\%, 10\%, and 1\%, respectively.
      The objects are color-coded according to the classification of \citet{zucker16}
{ with  the mid-IR active population in
blue and the mid-IR quiescent population in red to follow the optical
convention typically adopted in the literature.}
}
         \label{fig:mmol_vs_mstar}
   \end{figure}

Figure~\ref{fig:mmol_vs_mstar} shows the molecular gas mass as a function of stellar mass for the galaxies in our sample,
color-coded according to their class  determined in \citet{zucker16}.
A broad correlation exists between both quantities. 
{ Active galaxies have stellar masses between about log(\mstar/\msun) $\sim 9 -11,  $
whereas the mass range of quiescent galaxies is shifted to higher values 
(log(\mstar/\msun) $\sim 10 -11.5  $). The masses of canyon galaxies 
are in an intermediate range (log(\mstar/\msun) $\sim 9.5 -11.5  $).
With respect to the molecular gas }the distribution is remarkably stratified. Whereas for
active galaxies the molecular gas mass is mostly between 10 and 100\% of the  stellar mass
(with some objects having an even higher value), the molecular gas per stellar mass is considerably
lower for quiescent galaxies (between 1 and 10 \%). Canyon galaxies are right in between
these two groups, suggesting that they present a transition between them. 


   \begin{figure}
   \centering
\includegraphics[width=8cm]{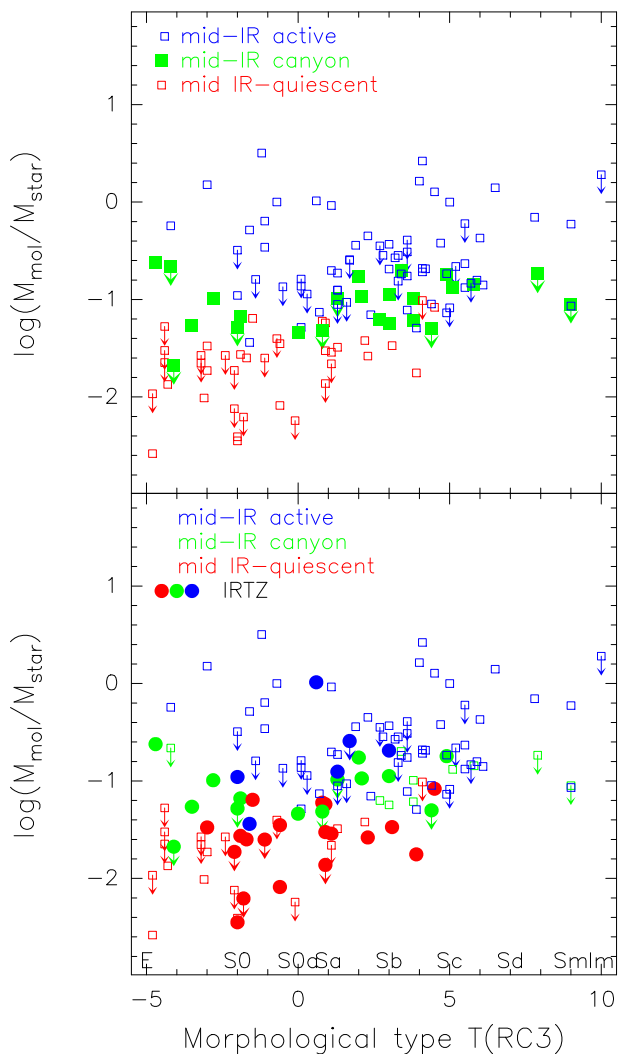}
      \caption{Ratio of molecular gas mass to stellar mass as a function of morphological type.
     The color coding is as in Fig.~1. 
       Filled circles (lower panel) denote galaxies belonging to the IRTZ . 
              }
   \label{fig:mmol_over_mstar_vs_morf}
   \end{figure}

Figure~\ref{fig:mmol_over_mstar_vs_morf} shows the ratio between molecular and stellar mass as
a function of morphological type. 
As expected, most active galaxies are late types and most quiescent galaxies are early types. 
Canyon galaxies are mostly late-type galaxies. 
%
We see again the distinct dichotomy in \mmol/\mstar\ between 
active and quiescent galaxies, where canyon galaxies have values within the lower end of the active
galaxies and span a similar range of morphological types.

In addition to distinguishing active, quiescent, and canyon galaxies, the lower panels also show
galaxies  belonging to the IRTZ. 
These have, as canyon galaxies, a lower value of \mmol/\mstar\ than active galaxies, but
are, to a larger extent,  early-type galaxies and many are classified as quiescent galaxies by \citet{zucker16}.
About half of the canyon galaxies also belong to the IRTZ showing that both criteria probe a 
similar class of objects.
%

The mean  { and median} values for the various groups,
calculated by taking into account the upper
limits in the calculation\footnote{
When calculating the mean values we take upper limits into account
with the program ASURV Rev, Astronomical Survival Analysis)
Rev, 1.1 \citep{lavalley92}, which is a generalized statistical package
that implements the methods presented by \citet{feigelson85}.}, 
are listed in
Table~\ref{tab:mean_mmol_over_mstar}. We list the values for the entire subsample and the values
for only late-type galaxies to compare the means for the same morphological group.
This is helpful to avoid mixing the effects of morphological type and activity class; early-type
galaxies are known 
to have a lower molecular gas content and SFR. 
The mean values confirm the visual impression from Fig.~\ref{fig:mmol_over_mstar_vs_morf}.
Canyon and IRTZ galaxies have a significantly ($\sim 5-7\sigma$)  lower \mmol/\mstar\  than active galaxies. 
This result holds when only considering late-type galaxies ($\sim 5\sigma$) .

\subsection{Molecular gas mass and star formation rate}

   \begin{figure}
   \centering
 \includegraphics[width=8cm,angle=270]{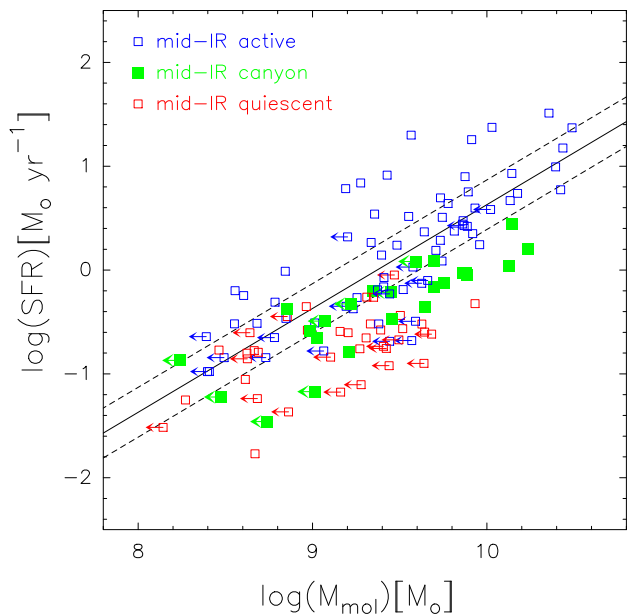}
     \caption{Star formation rate vs. total molecular gas mass, both calculated as explained in Sect. 2. 
     The color coding is as in Fig.~1. 
      The lines show the
      mean value (full line) and dispersion (dashed lines) found by \citet{bigiel11} for a sample of spiral and starburst galaxies. 
              }
         \label{fig:sfr_vs_mmol}
   \end{figure}

Figure~\ref{fig:sfr_vs_mmol} shows a comparison of the SFR as  a function of molecular gas mass for active,
quiescent, and canyon galaxies.
We include 
the result (best-fit and standard deviation) of  \citet{bigiel11}  who found a linear relation
between the SFR and molecular gas mass for a large sample of spiral and starburst galaxies.
Active galaxies roughly follow this relation, albeit with a large scatter and a small offset ($\sim 0.1 - 0.2$ dex).
{ Quiescent galaxies have a pronounced offset toward lower SFR for a given molecular gas mass and
they lack the high molecular gas masses present in some active galaxies; the upper limit in log(\mmol/\msun) is $\approx$
9.8 for quiescent galaxies, whereas for active galaxies it is  $\approx$ 10.5.}
 Canyon galaxies
have a considerably lower SFR for a given molecular gas mass than active galaxies, in the same range as the
quiescent galaxies. 

\begin{table}
\caption{Mean { and median log(SFE) (in units of yr$^{-1}$}) for different samples}
\begin{tabular}{lccc}
\hline
Class &mean & median & n/n$_{\rm up}$\tablefootmark{a} \\
\hline
{Full sample} & & \\
Active    & -9.17 $\pm$ 0.05 & -9.25 & 68/17\\
Canyon  &  -9.65 $\pm$ 0.07 & -9.71 & 24/8 \\
IRTZ & -9.58 $\pm$ 0.06 & -9.59 & 38/10 \\
quiescent  &  -9.67 $\pm$ 0.06 & -9.61 & 38/16  \\
\hline
{Late-type galaxies ($T > 0$)} & & \\
Active    & -9.26$\pm$ 0.05  & -9.27 & 52/12\\
Canyon  &  -9.80 $\pm$ 0.05 & -9.89 & 16/5 \\
IRTZ & -9.71 $\pm$ 0.07 & -9.85 & 20/5 \\
quiescent  &  -9.87 $\pm$ 0.09 & -9.96 & 13/3 \\
\hline 
\end{tabular}
\tablefoot{
\tablefoottext{a} {Total number of galaxies (n) and number of upper limits (n$_{\rm up}$)}.
}
\label{tab:mean_sfe}
\end{table}

   \begin{figure}
   \centering
 \includegraphics[width=8cm]{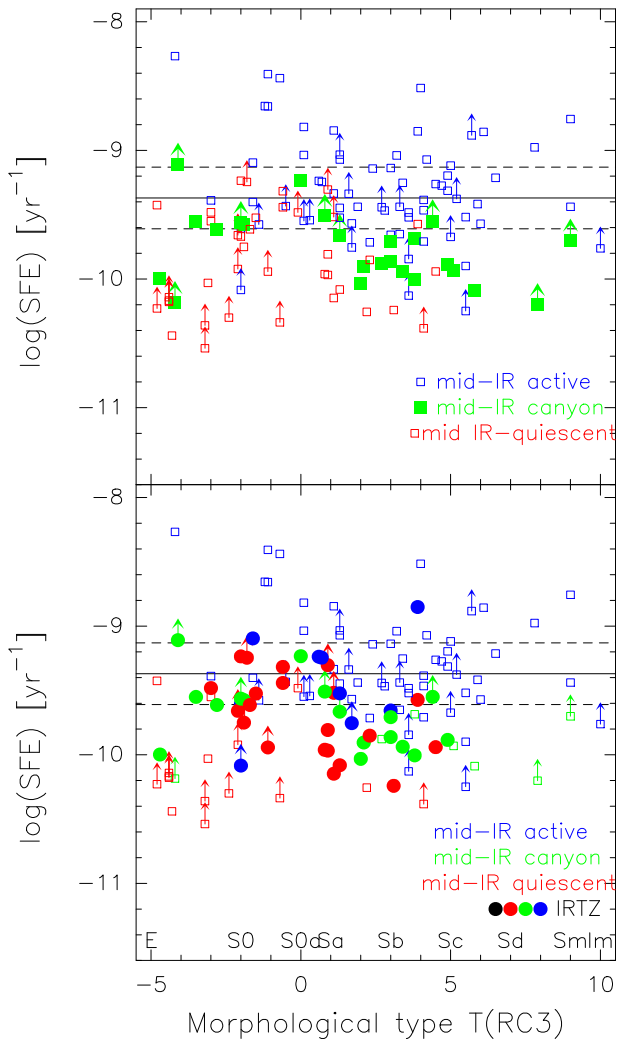}
      \caption{SFE  as a function of morphological type.
     The color coding is as in Fig.~1. 
       Filled circles denote galaxies belonging to the IRTZ (lower panel). 
      The lines show the
      mean value (full line) and dispersion (dashed lines) found by \citet{bigiel11} for a sample of spiral and starburst galaxies. 
                  }
         \label{fig:sfe_vs_morf}
   \end{figure}

Fig.~\ref{fig:sfe_vs_morf} shows the distribution of the SFE as a function of the morphological types. 
The mean { and median} values for the three groups
are listed in Table~\ref{tab:mean_sfe}.

The mean value for active galaxies lies slightly higher, but within the scatter of  the 
mean value found by \citet{bigiel11} (-9.37$\pm$0.23).
Canyon and IRTZ
galaxies have a clearly lower SFE than active galaxies ($> 5 \sigma$), both for the entire subsample and for late-type galaxies alone.
Galaxies belonging to the IRTZ have very similar values for the SFE as the canyon galaxies.
Galaxies classified as quiescent outside the IRTZ are mostly early types and have very low SFE and a low CO detection rate.

For late-type galaxies, all the canyon/IRTZ  galaxies lie on the
low end of the distribution of the SFE. Toward earlier types (Sa and earlier), the SFE of canyon/IRTZ becomes similar to
the values of quiescent galaxies. { This is in principle a surprising result, however, it should be taken with caution mainly 
because of the low SFRs of early-type galaxies. Any SFR indicator is less reliable for these low values \citep{leroy12} and,
in particular,  the mid-IR dust emission 
because of the contribution of asymptotic giant branch stars to the mid-IR \citep{kaviraj07}.
\citet{lenkic16} estimated the effect of this contamination for a sample of CG galaxies and found
that for about 40\% of quiescent galaxies the contribution of AGB stars to the 24~\mi\ emission could 
be 50\% or more whereas for active the contribution of AGB stars was negligible.
A comparision with a different SFR tracer would be desirable. We carry out this comparison
in Sect.~\ref{comparison_sfr_tracer}, but the number of canyon/IRTZ early-type galaxies
available for this test is too low to draw any useful conclusions.

%
%

\subsection{Overall picture}

We have found that canyon/IRTZ galaxies  have a molecular gas fraction (\mmol/\mstar) in between 
active and quiescent galaxies and a SFE as low as quiescent galaxies. This suggests that canyon/IRTZ galaxies
are indeed in transition between the active and quiescent phase. 
Their SFE is already as low as for quiescent galaxies, which might indicate similar physical properties
of the molecular gas, but they still have a larger molecular gas content. We speculate that
their future, in the absence of  additional external influences, is to continue forming stars with
a low efficiency and thereby continuously decreasing their molecular gas content. 
}

\section{Discussion}

\subsection{Comparison of SFR and \mstar\ to other studies}
\label{comparison_sfr_tracer}

We used a prescription to calculate the SFR and stellar mass entirely based on WISE
data. In this way, we managed to maximize the sample size.
As a test of the robustness of our results, we compared the values for the
SFR and \mstar\ with those of \citet{bitsakis11} and \citet{lenkic16} and repeated our analysis with
their values for those objects that are in common.

\citet{bitsakis11} carried out a multiwavelength analysis, based on data from the UV
to near-infrared 
for a sample  of 135 galaxies and additional mid- and far-IR from \spitzer\ and  AKARI for a subsample. 
They used the model of \citet{dacunha08} to fit the SED and derive stellar mass, SFR, and extinction.
We compared the values of \mstar\ and the SFR for those galaxies that were in common with our sample
(112 objects in total, 84 objects detected in all WISE bands, and 66 galaxies detected by WISE and with  CO data).
Both the SFRs and stellar masses obtained by \citet{bitsakis11} correlate well and linearly with our values
(correlation coefficients $r= 0.67$ and 0.91 and bisector slopes $1.05\pm0.05$ and $0.99\pm0.13$, respectively).
There is a small, constant  offset between our SFR and \mstar\ and those derived by \citet{bitsakis11};
our \mstar\  is about 50\% lower than that of  \citet{bitsakis11} and our SFR is about 30\% higher.

We reanalyzed  for this smaller sample, consisting of objects in common with  \citet{bitsakis11},  
using the SFR and \mstar\ derived by  \citet{bitsakis11}.
The results 
 confirmed the trends found with the larger, WISE-based sample with very similar values both
 for the SFE and molecular gas fraction. Thus, also for this smaller sample, the mean log(SFE/yr$^{-1}$) for   canyon galaxies
 ($-9.88 \pm0.09$ for eight objects) and IRTZ galaxies ($-9.89 \pm0.14$ for 14 objects) is significantly below the
 value for active galaxies ($-9.14 \pm0.14$ for 40 objects), and the molecular gas fraction log(\mmol/\mstar)
 is lower in canyon  ($-1.26 \pm0.16$) and IRTZ galaxies ($-1.79 \pm0.12$) than in active galaxies ($-0.93 \pm0.7$).

A different way to calculate the SFR in CG galaxies was employed by  \cite{lenkic16}. They derived the SFR for a sample 
of 175 galaxies based on \spitzer\ MIPS data and UV data from
{\it Swift} UVOT. Their derivation of the SFR is thus sensitive to both the dust-enshrouded and unextincted SF.
{
There exists a good correlation between the \citet{lenkic16} SFR and our SFR 
(correlation coefficient $r= 0.87$) for the 53 galaxies that are in common
in both samples,  have both UV and Spitzer
24~\mi\ data and are detected in all  WISE bands. The slope is, however, not unity, and at low SFRs the
values of  \citet{lenkic16} are about a factor of  $\sim 3$ higher
than those of \citet{zucker16}, indicating that the UV emission is, especially at low SFRs,  important
to trace the total SFR.

We reanalyzed the subsample of 42 galaxies that are in common and have  CO data 
using the (UV+24~\mi) SFR  from  \citet{lenkic16}.
The results were inconclusive. 
At high molecular gas masses (above
$\sim 10^9$ \msun)  the SFE of canyon/IRTZ (3 objects in each group) is $0.5-1.0$~dex below the
value of active galaxies, consistent with our results. However, at lower masses,
the 3 canyon and 4 IRTZ have SFEs in the range of  active galaxies.
A larger sample of HCG galaxies with  UV data, especially at low masses,
is needed in order to perform a statistically significant test.


}

\subsection{Role of molecular gas in  galaxy transition}

Our analysis has shown that the molecular gas is  a crucial factor in the transition of galaxies from
actively star forming to quiescent. Both the molecular gas fraction, \mmol/\mstar, and the
SFE decrease significantly in the  transition phase (canyon/IRTZ galaxies).
This means that in the IR transition phase   the molecular 
gas content becomes lower and the remaining molecular gas forms stars less efficiently. 
{ In the following we discuss possible causes for these findings.}

%

A loss of atomic gas is a well-known process in CGs
with deficiencies reaching very high values (more than 90\%) in some cases \citep{verdes-montenegro01}.
However, the molecular gas does not seem to be strongly affected by this. Previous surveys of the
molecular gas \citep{leon98,verdes-montenegro98}  found no strong decrease in the molecular gas compared to field galaxies. 
\citet{martinez12} compared molecular and atomic gas deficiencies in a sample of HCG late-type galaxies and found no
significant correlation; even galaxies that were strongly deficient  in HI can show a normal \htwo\ content.
These authors explained this different behavior with the more extended distribution of HI compared to \htwo .
This means that overall there is no 
lack of molecular gas in late-type galaxies in HCG and that this deficiency is only present for IR transitioning galaxies.

The lack of molecular gas could either be due to a decrease in the formation of \htwo , the destruction of the 
molecular gas, or a loss
of the molecular gas, for example, by tidal  forces, in a similar way as HI. So far it is unclear what the relevant process is.
We can only exclude a lack of atomic gas  supply as a reason because
observationally there is no relation between a deficiency in HI and \htwo. The formation rate of the \htwo\ is proportional to the gas density, which might be lower in
the transition phase. A low gas density would at the same time explain the low SFE (see below). In the transition
 phase tidal forces might also be strong enough
to affect the more central areas of the galaxies where molecular gas is present. If tidal forces were the relevant process, we would expect to observe molecular gas outside the
galaxies or in tidal arms or streams. Thus, further observations of the molecular gas distribution and kinematics are necessary
to distinguish between the different processes.


The low SFE of canyon and IRTZ galaxies { shows that 
a lack of molecular gas  is not the only
reason for a decrease in SF during the transition. Instead,  
the remaining molecular gas 
has  lost its ability to form stars.}
A possible reason could be 
a perturbation of the molecular gas, possibly due { to turbulence injected by}  shocks
that are produced by the interaction with neighboring galaxies or intra-group gas.
This has been seen in the case of HCG 57a, which is a galaxy in close interaction with HCG 56d, where interferometric CO data has
shown the presence of perturbed molecular gas that might explain its low SFE
 \citep{alatalo14a}.  Collision with the IGM has been suggested by \citet{cluver13} to 
  cause enhanced warm \htwo\ emission. The best-studied example for this kind of process 
 is HCG~92 (Stephan's Quintet), where
shocks and turbulence are strongly suggested to be responsible for
the suppression of SF in the intra-group gas  \citep{cluver10,appleton13,appleton17}. 
Furthermore, \citet{bitsakis16} found indications for the presence of shocks in CG galaxies
below the SF main sequence.


Our results allow us to provide additional pieces of evidence to understand the evolution of galaxies in 
CGs. Altogether, the following evolutionary sequence seems to be likely
 \citep[see also][who suggested similar models]{verdes-montenegro01,walker10,walker12,cluver13,alatalo14b, bitsakis16}.
In an early stage,  the atomic gas of galaxies in CGs gets extracted from the galaxies from tidal forces  and 
gets distributed in the intra-group medium \citep{verdes-montenegro01,borthakur10}.
The outer parts of galaxies are affected first, such that the molecular gas and dust \citep{bitsakis14} are
not (much) depleted
in this phase. This explains the lack of molecular gas deficiency in CG galaxies in general.
As CGs evolve, the velocity dispersion increases  and interactions, such as tidal interactions between galaxies and interactions
with the intra-group gas, become more important.
At this stage, the molecular gas content and SFE become affected.

Although in this scenario the decrease of the molecular gas is not causally related to a lack of atomic gas, there is a 
temporal sequence and we would expect transitioning galaxies with a low molecular gas content to be 
deficient in HI as well. The necessary high-resolution HI data are available for 21 transitioning galaxies 
\cite[3 canyon and 18 IRTZ galaxies; see Tab. 5 in ][]{martinez12} and confirm our expectation.
Fourteen galaxies have 10\% or less of the expected HI mass,
six galaxies have between 10 and 30\%,  and only one  galaxies has  a slight deficiency (60\% of the expected HI).




{ 
\subsection{Relevance for galaxy evolution}

%

The cosmic evolution of galaxies has been characterized by a strong decrease in the SFR since about 
$z \sim 1-1.5$  \citep{madau98,hopkins06};  the reason for this is still an open question.
%
%
There has been a large effort in recent years to better understand the cosmic SF history
with different approaches. One way is to study the so-called galaxy main sequence (MS),
which is the
 relation between the SFR and stellar mass  followed by 
star-forming galaxies \citep{noeske07}.
The slope of this relation gives information
about how the specific SFR, sSFR = SFR/\mstar, changes as a function of 
stellar mass.  There are indications that the slope decreases at high mass,
i.e., that the sSFR is lower for massive galaxies, thus implying a quenching of SF at high
masses.
The turnover mass increases from $\sim 10^{10}$ \msun\ at $z \sim 0$ \citep{brinchmann04}
to $\sim 5 \cdot 10^{10}$ at $z \sim 1$
\citep{schreiber16}.

In order to understand the possible causes of this quenching and cosmic SF history
in general, measurements of the gas mass are necessary and it is useful to quantify
the SFE and the (molecular or total) gas fraction  as a function of redshift. Since it is observationally difficult
to measure the gas content for distant galaxies, not many studies 
have been performed thus far.
\citet{combes13} observed a sample of 39 ultra-luminous IR galaxies (ULIRGs) with $z \sim 0-1$ 
in CO(1-0) and found a decrease of both the SFE and the molecular gas fraction for decreasing $z$. These authors
concluded that both factors seems to be relevant for the decrease of the cosmic
SF history. 
\citet{schreiber16} derived the total gas mass
 from HERSCHEL dust data for 
 a sample of galaxies in the CANDELS field  (at z$\sim 1$) and found  that for  \mstar $\gtrsim 5\cdot 10^{10}$ 
both the sSFR and the SFE decreases.
 The total gas fraction showed no similar decline, indicating
that a lack of gas is not the reason for the low sSFR. 
These authors suggested that the decrease in SFE implies a slow
quenching (the cause of which is unclear) such that  galaxies decrease their
SFR slowly while remaining on the MS, instead of suffering a catastrophic event like a merger.
A decrease in the SFE  with stellar mass was also found by \citet{saintonge11b}
for local, mass-selected galaxies and a weak decrease in the molecular gas fraction was found by  \citet{saintonge11a}.


When we compare these results to ours, we need to keep in mind that the
CG galaxies  live in a special environment
where interactions, both between galaxies and with the intra-group medium, play a crucial role.
In spite of the interaction, the galaxies in our sample are, however, not in a starburst mode
of SF; instead they form stars in a normal way, similar to spiral galaxies.
Because of this, the mean SFE of the active group is similar to the value of \citet{bigiel11} for 
spiral galaxies. Extreme starbursts have much higher SFEs  \citep[e.g.,][]{daddi10}.
Similarly, the majority of actively star-forming CG galaxies in the study of \citet{lenkic16} was consistent with the
local MS in the parameter space of SFR versus \mstar\ found by \citet{chang15}.
Furthermore, their  molecular gas surface density ($\lesssim 50$ \msun~pc$^2$),
is much lower than the surface density of $\gtrsim 100$ \msun~pc$^2$ typical for starbursts.

All these studies identified a decrease of the SFE as a reason for 
the quenching of SF. It is interesting that this result holds for very different sample,
from ULIRGs which form stars in a starburst mode, to normal galaxies.
For our sample we speculate that the environment is responsible
for the decrease in SFE. It is still unclear, but important to find out, whether the decrease of SFE in the other samples has the same
origin. The finding by \citet{schreiber16} that the decrease in the gas fraction is not related to the quenching of SF  contrasts with our results and is most likely due to  the 
difference in environment, in particular the dense, but low-mass environment of our sample.

}

\section{Conclusions and summary}

We analyzed the molecular gas content  based on 
CO(1-0) data from the literature and  from our own new data IRAM 30 m observations 
for a sample of { 130 } galaxies in CGs with detections from the
\wise\ satellite in all four bands. The  mid-IR \wise\ colors were used by \citet{zucker16} to classify these galaxies
into actively star-forming, quiescent, and canyon (i.e., transitioning between both phases) galaxies.
We  used these data and  classifications  to compare the SFE and molecular gas fraction, \mmol/\mstar,
between the active, quiescent, and transitioning galaxies.
We considered two types of transitioning galaxies: Canyon \citep{zucker16} and IRTZ galaxies \citep{alatalo14b}, selected based on the
WISE [W2-W3] color.

We found a significantly lower molecular gas fraction for canyon and IRTZ galaxies { than in  active galaxies, but still higher
than in  quiescent galaxies}. This 
indicates that the transition from active to quiescent goes along with a decrease in molecular gas { and that
canyon/IRTZ galaxies are indeed in a transition process.}
The reason for this decrease is not entirely clear. A lack of gas supply can be  excluded because the deficiencies in
HI and \htwo\ do not correlate \citep{martinez12}. Possible causes are a decrease in the gas density (e.g., due to a perturbation  of the gas), making
molecular gas formation less efficient, or the tidal removal of molecular gas.

We  found a significantly lower SFE for canyon and IRTZ galaxies { than for active galaxies, which is comparable
to the mean value of quiescent galaxies.}
This shows that the remaining molecular gas has lost its ability to form stars efficiently in the transition.
A perturbation of the molecular gas, possibly from  shocks produced by the interaction with
neighboring galaxies or intra-group gas is a likely reason.


Thus, we found that molecular gas data can provide a crucial clue to the conditions of environmentally
driven galaxy evolution in CGs. Based on these new data,
we propose a possible scenario in line and extending conclusions from previous studies
 \citep[e.g.,][]{verdes-montenegro01,walker10,walker12,cluver13,alatalo14b, bitsakis16}.
for the evolution of galaxies in HCGs. In a first phase, galaxies lose their atomic gas, sometimes to a dramatic extent.
Eventually, when the galaxies are in the canyon/IRTZ phase the molecular content and SFE also decrease. In this phase,
the ISM is strongly affected, and this is most likely due to interaction with neighboring galaxies and/or the intra-group environment.

\begin{acknowledgements}
We appreciate very much the useful thoughts and comments from the referee.
UL acknowledges support by the research projects
AYA2014-53506-P from the Spanish Ministerio de Econom\'\i a y Competitividad,
from the European Regional Development Funds (FEDER)
and the Junta de Andaluc\'ia (Spain) grants FQM108.
We acknowledge the usage of the HyperLeda database (http://leda.univ-lyon1.fr) and of the
Nasa Extragalactic Database (NED, https://ned.ipac.caltech.edu).
This work is based on observations carried out under project number  167-16 with the IRAM 30 m telescope.
IRAM is supported by INSU/CNRS (France), MPG (Germany), and IGN (Spain). We would like to thank the 
IRAM staff at the 30m telescope warmly for their support during the observations.
\end{acknowledgements}


\bibliographystyle{aa}
\bibliography{biblio_wise}

\appendix

\section{Figures of spectra}


\begin{figure*}
\centerline{
\includegraphics[width=3.6cm]{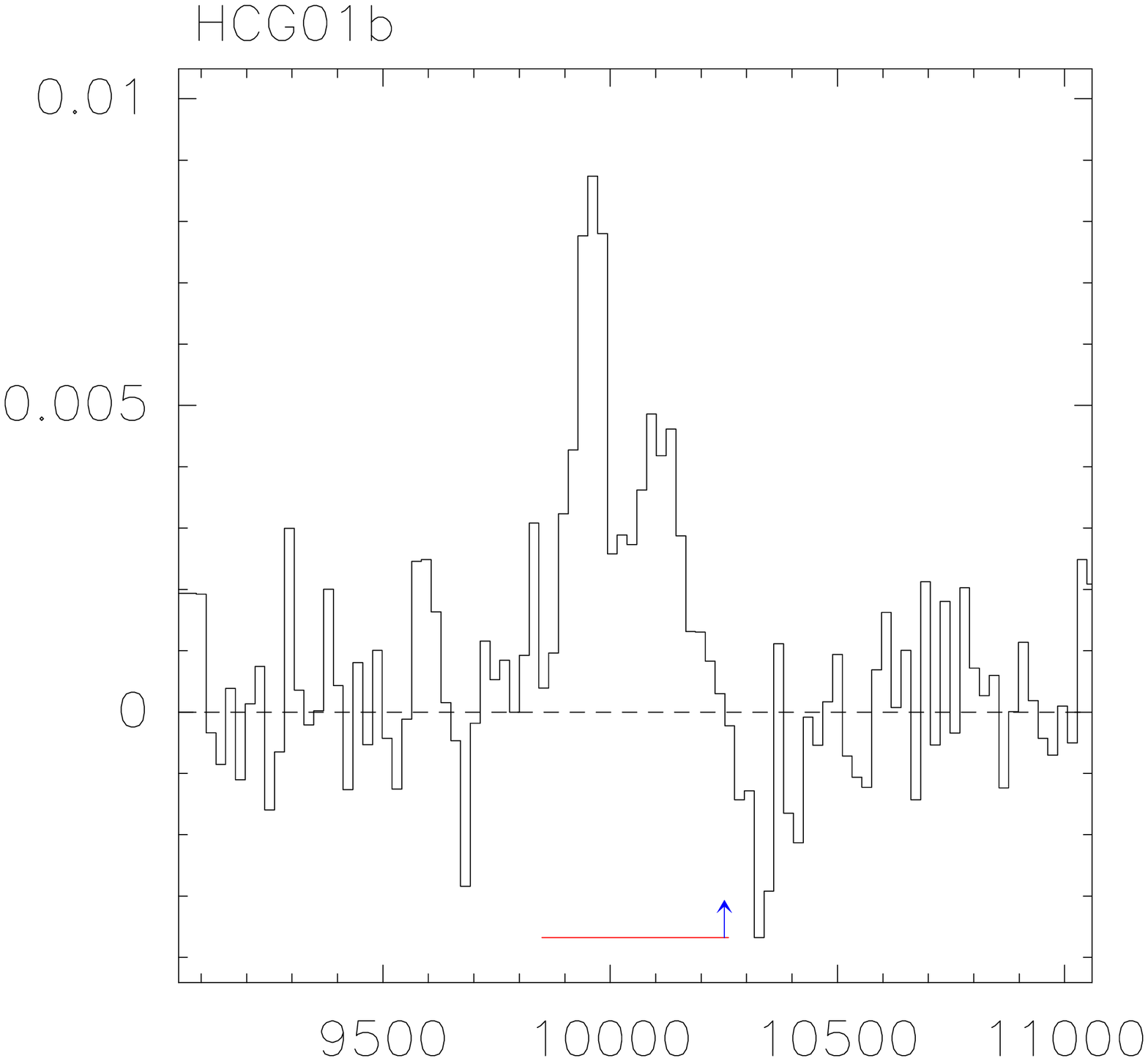}
\hspace{0.1cm}
\includegraphics[width=3.6cm]{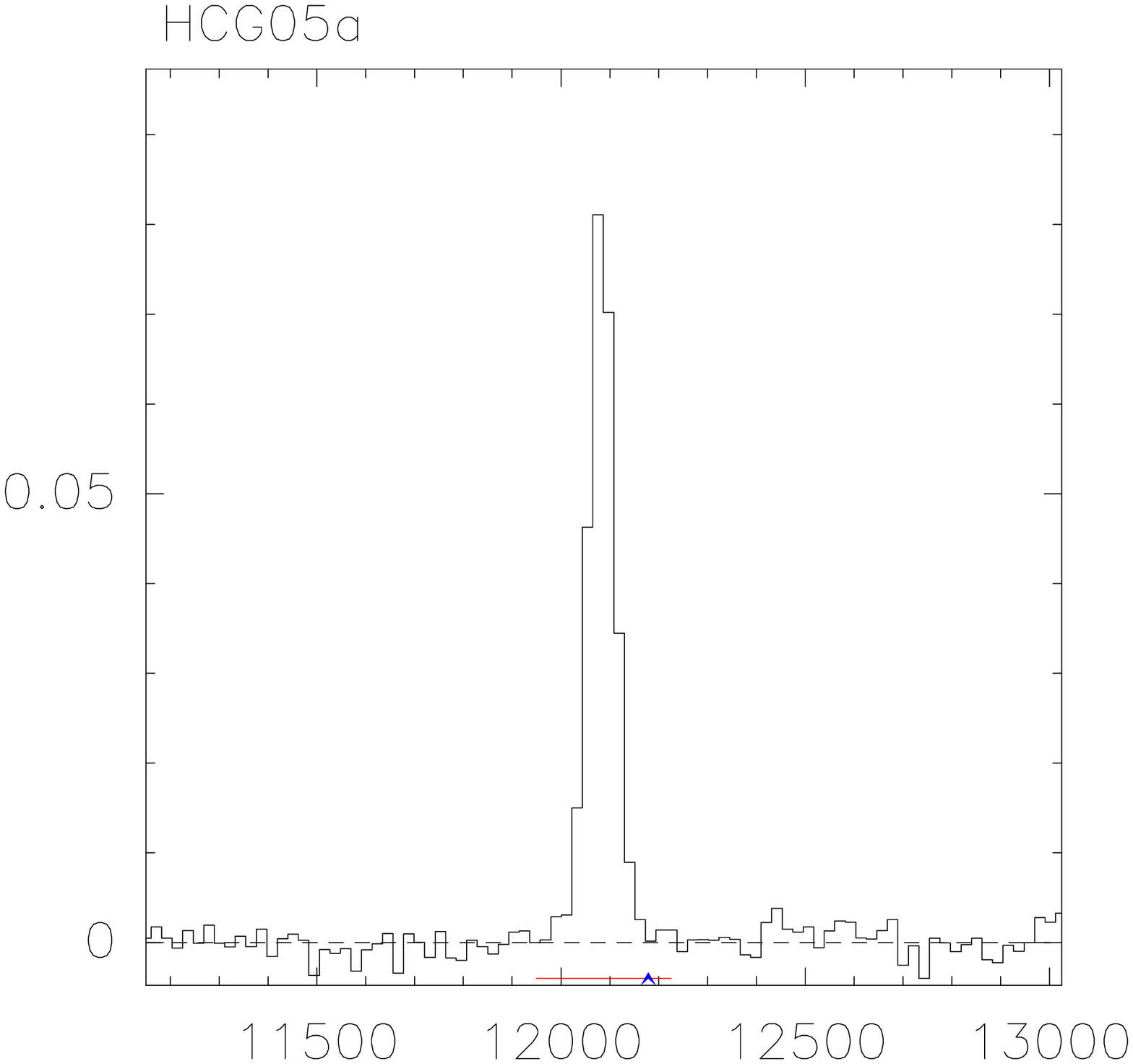}
\hspace{0.1cm}
\includegraphics[width=3.6cm]{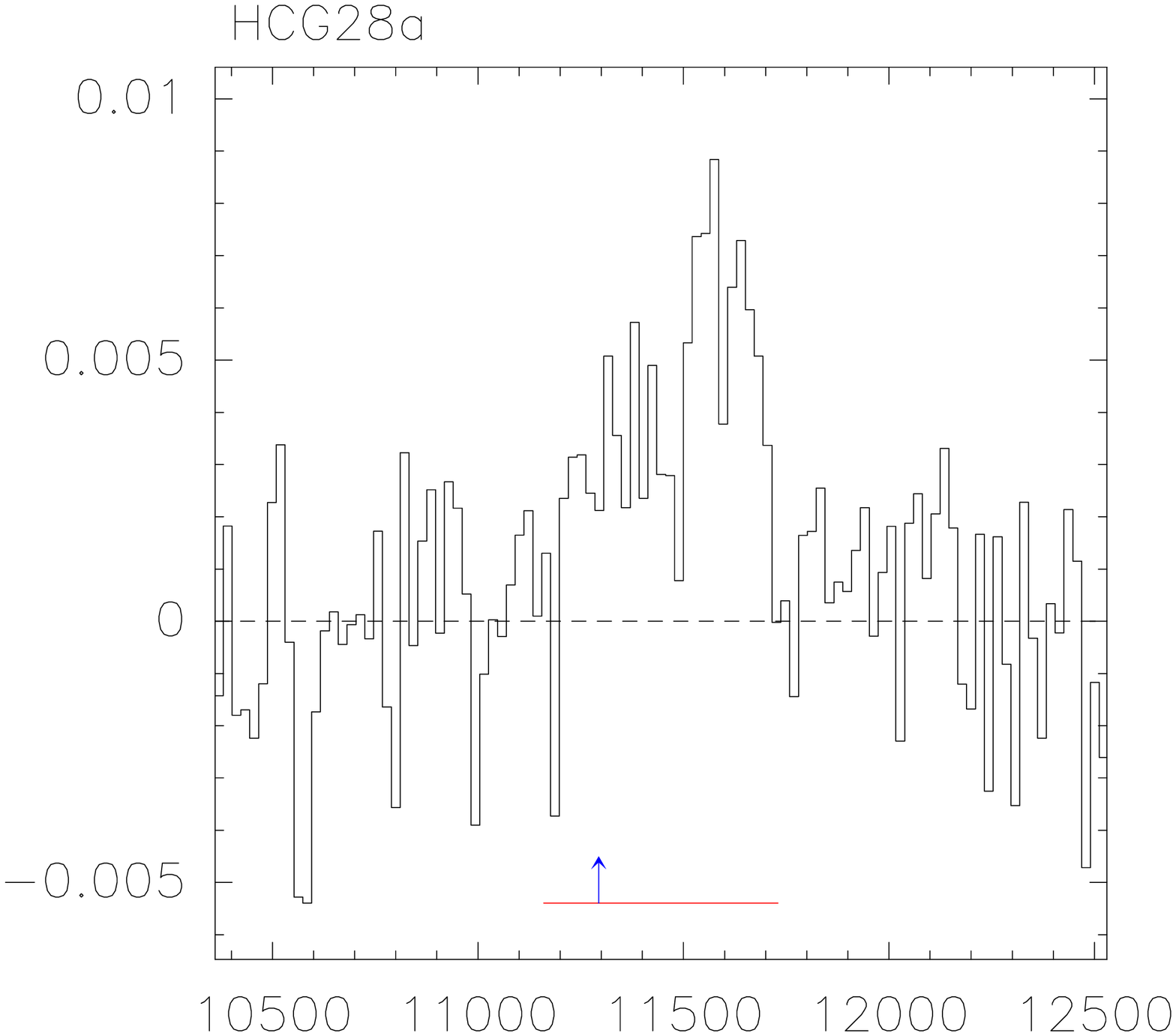}
\hspace{0.1cm}
\includegraphics[width=3.6cm]{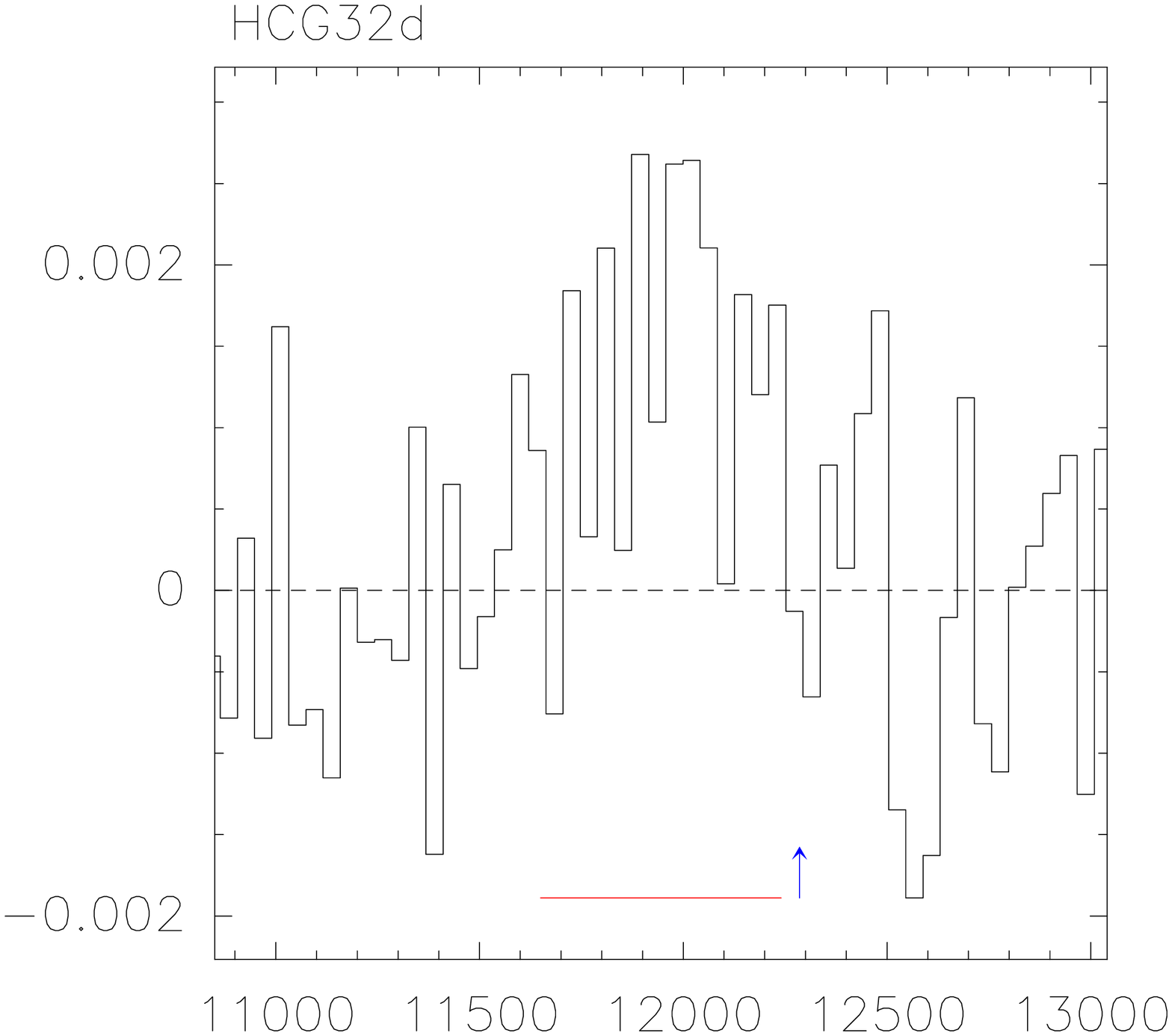}
}
\vspace{0.3cm}
\centerline{
\includegraphics[width=3.6cm]{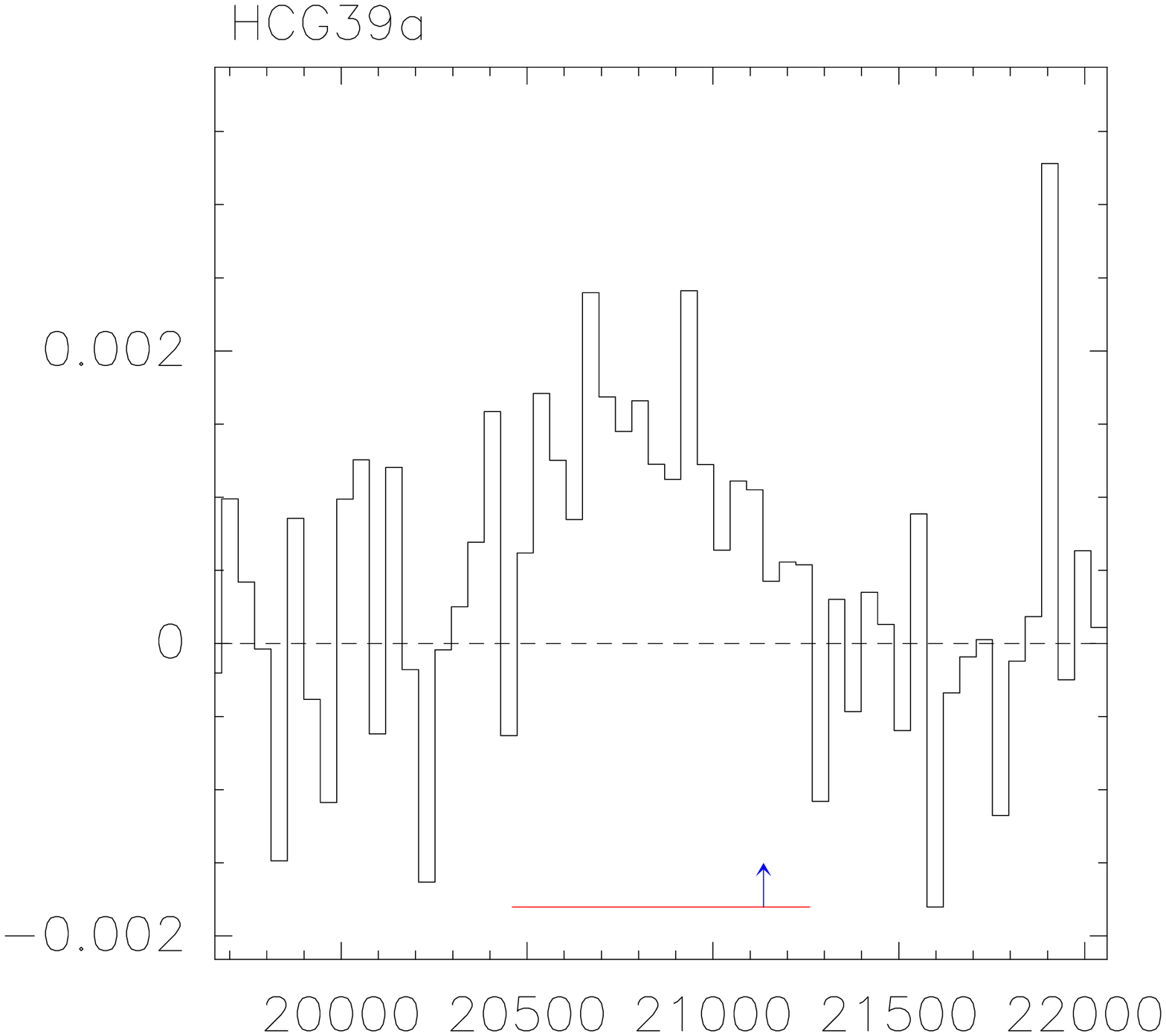}
\hspace{0.1cm}
\includegraphics[width=3.6cm]{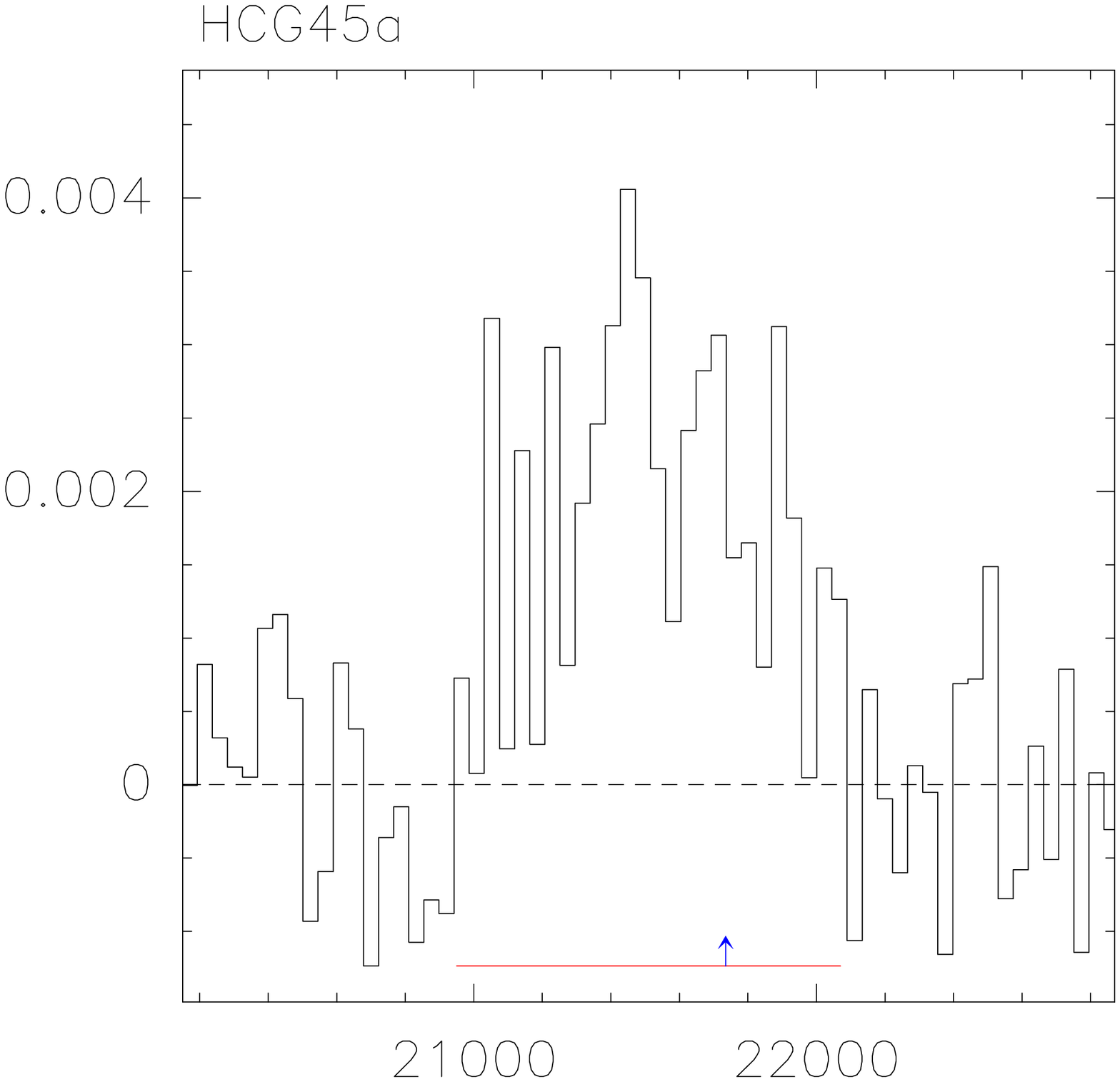}
\hspace{0.1cm}
\includegraphics[width=3.6cm]{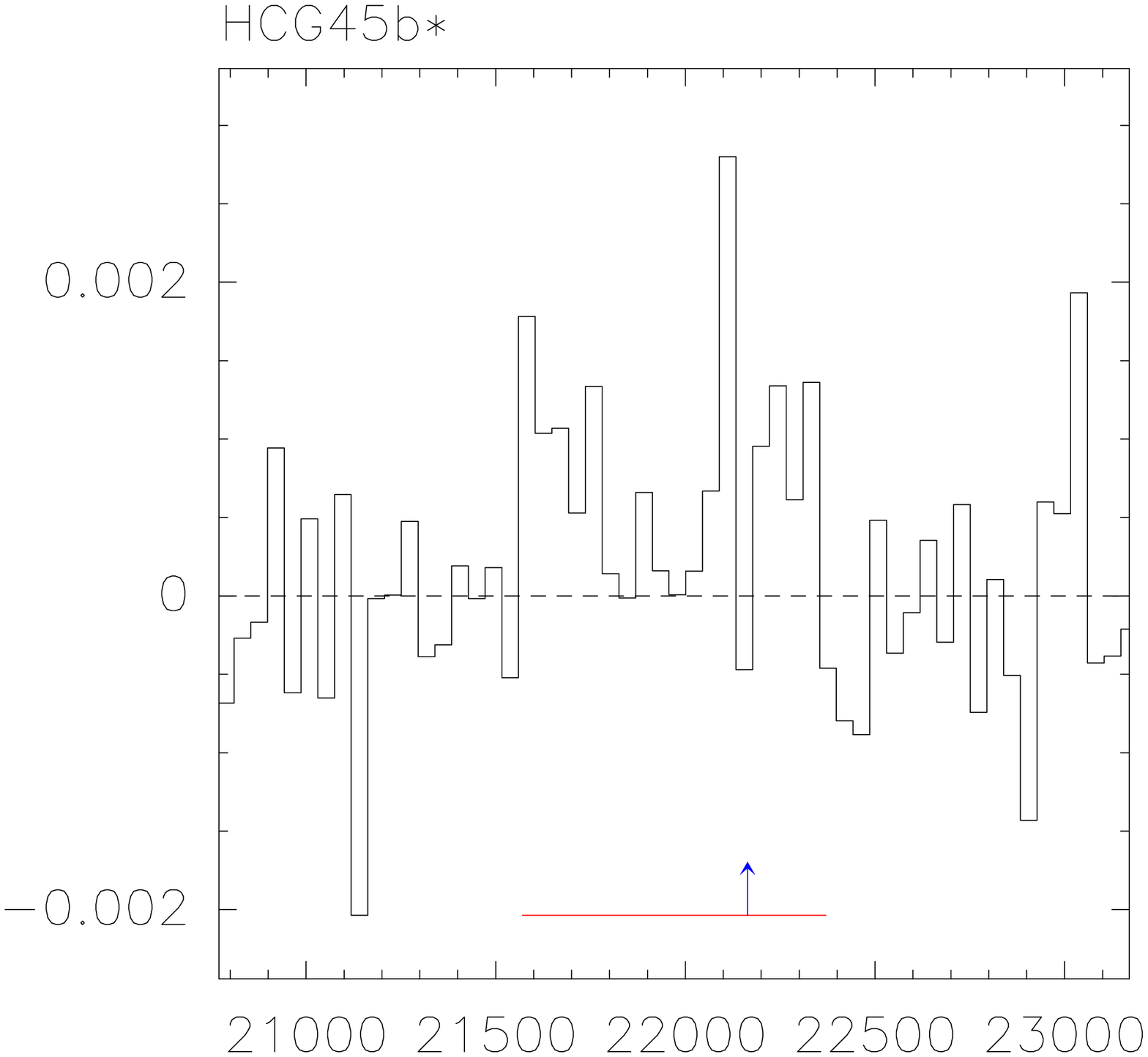}
\hspace{0.1cm}
\includegraphics[width=3.6cm]{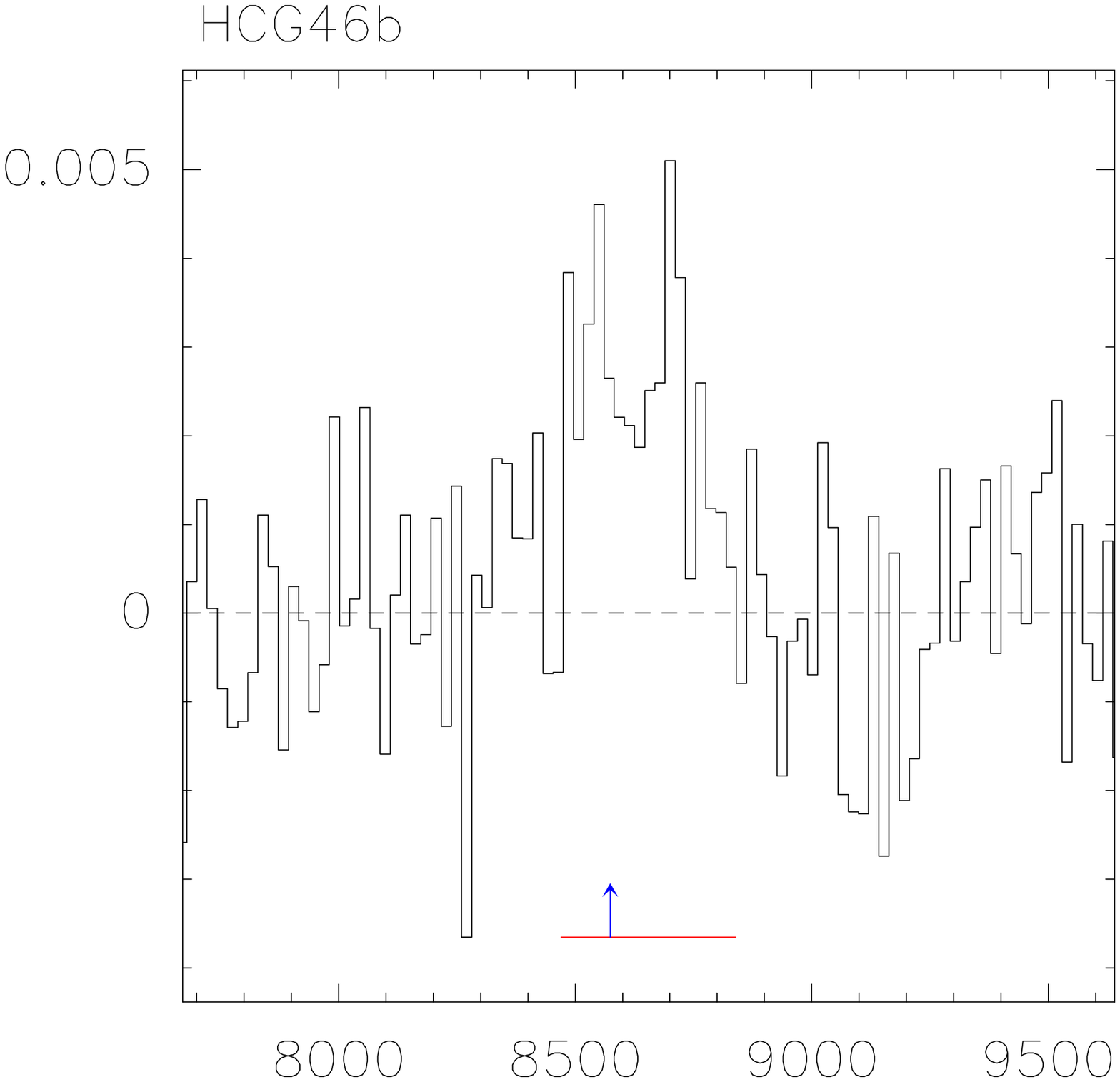}
}
\vspace{0.3cm}
\centerline{
\includegraphics[width=3.6cm]{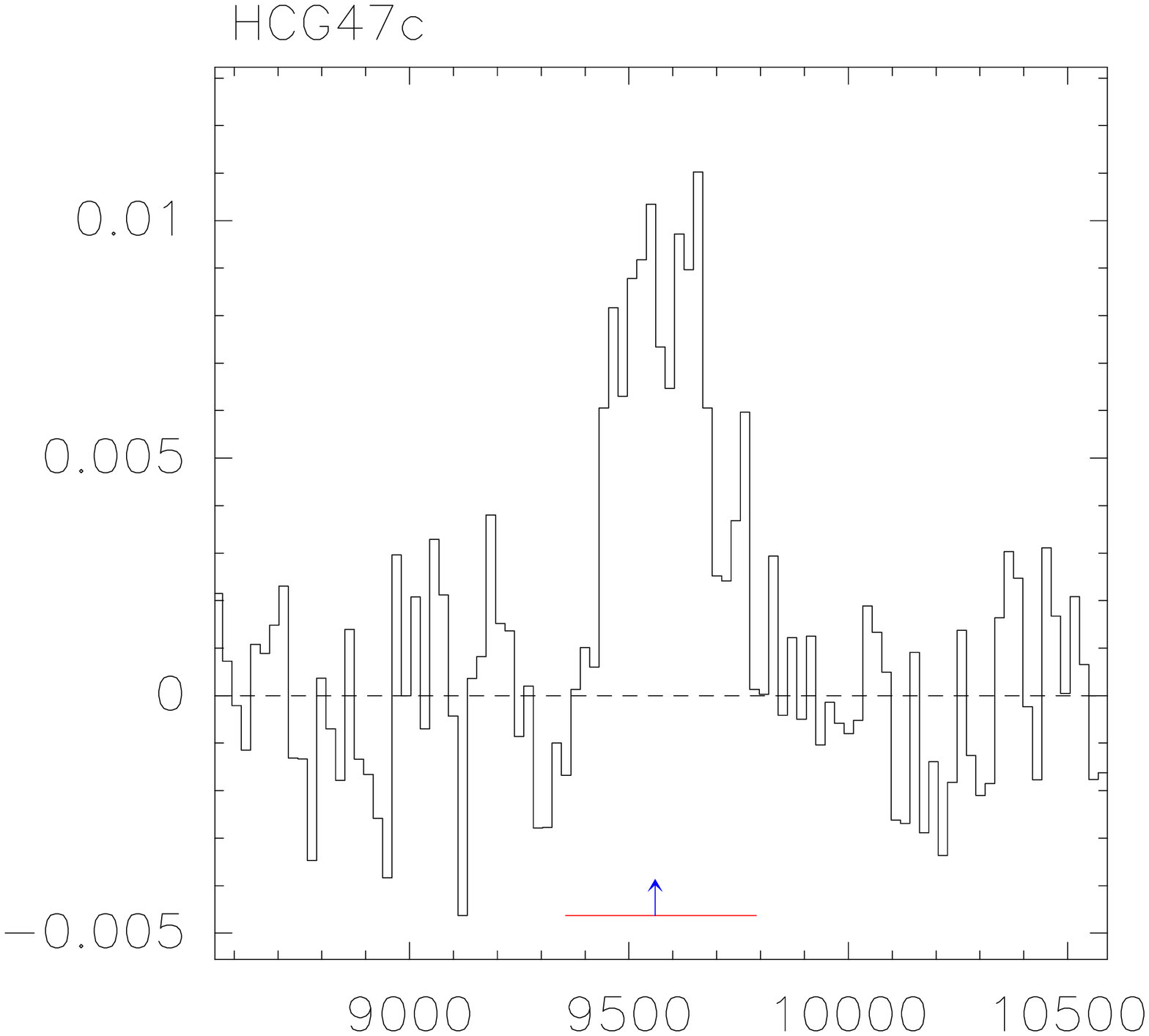}
\hspace{0.1cm}
\includegraphics[width=3.6cm]{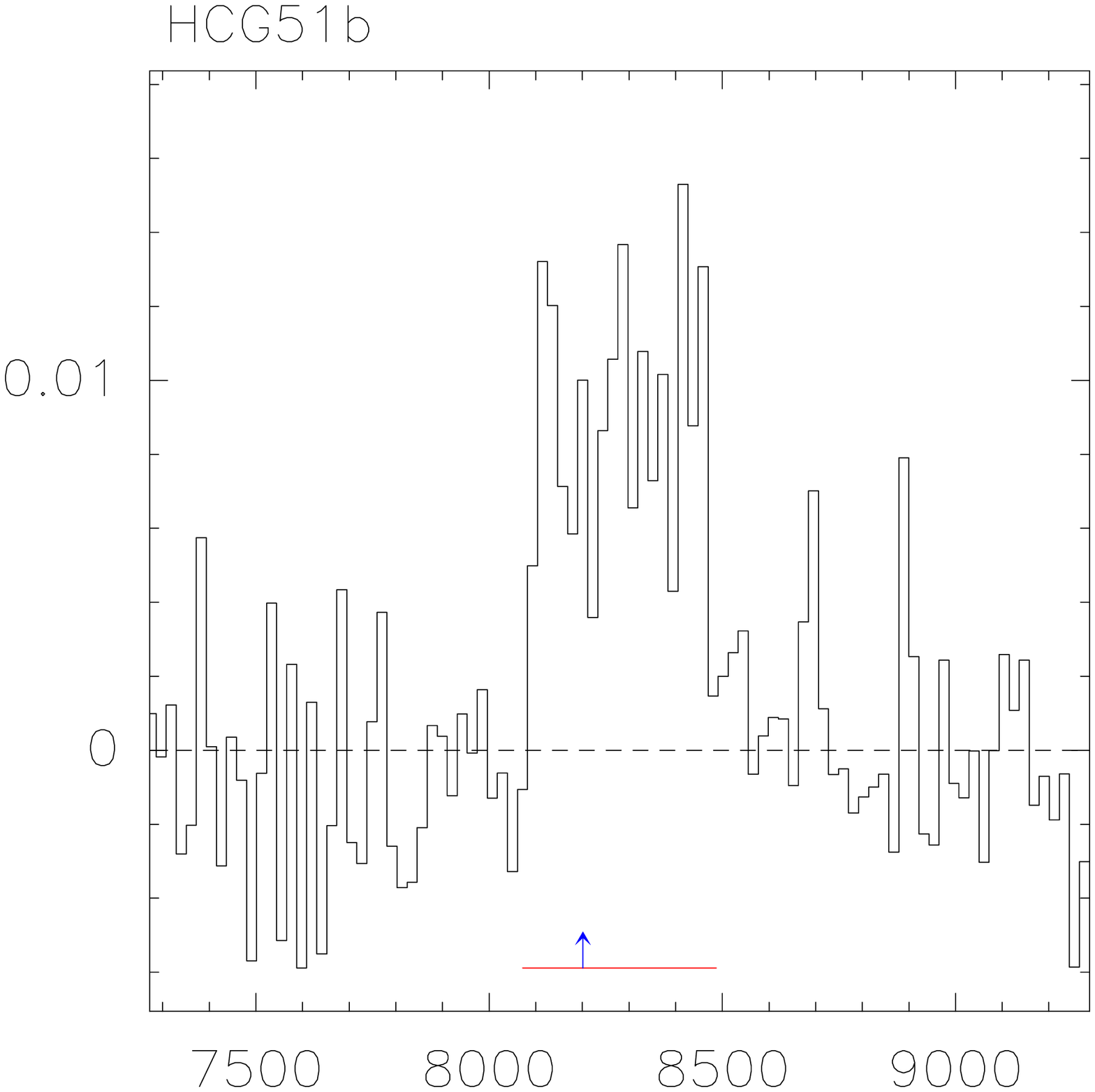}
\hspace{0.1cm}
\includegraphics[width=3.6cm]{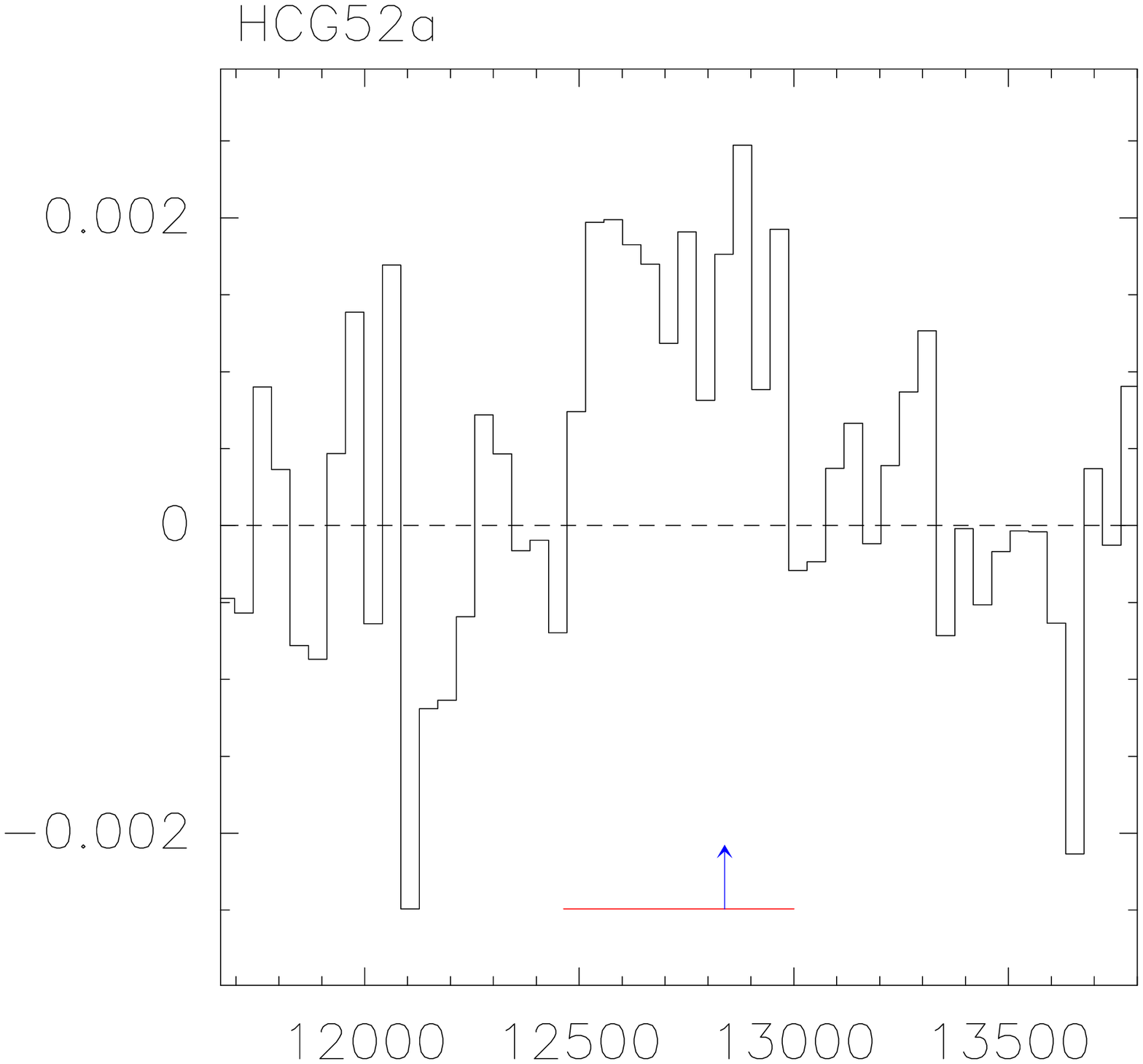}
\hspace{0.1cm}
\includegraphics[width=3.6cm]{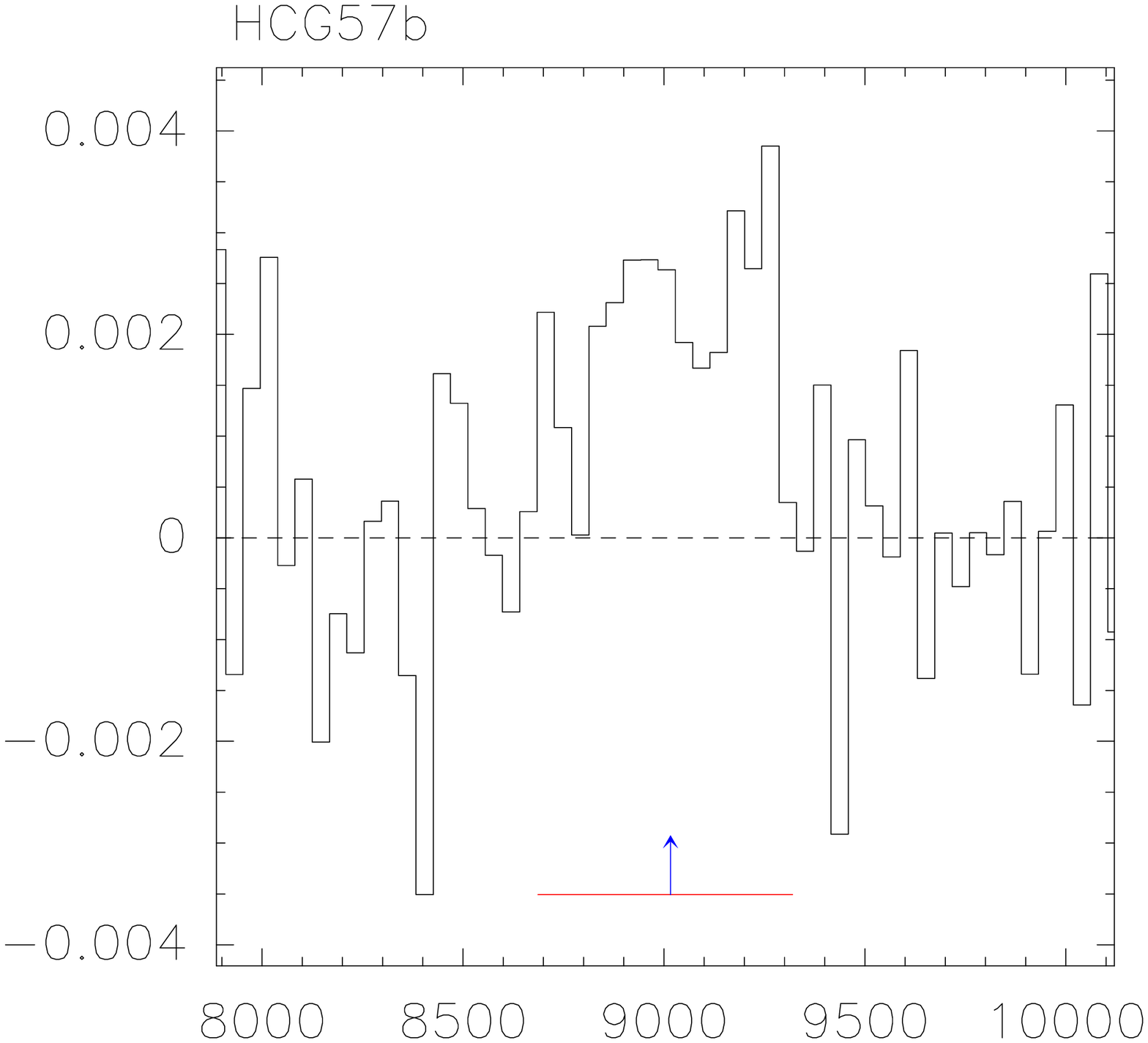}
}
\vspace{0.3cm}
\centerline{
\includegraphics[width=3.6cm]{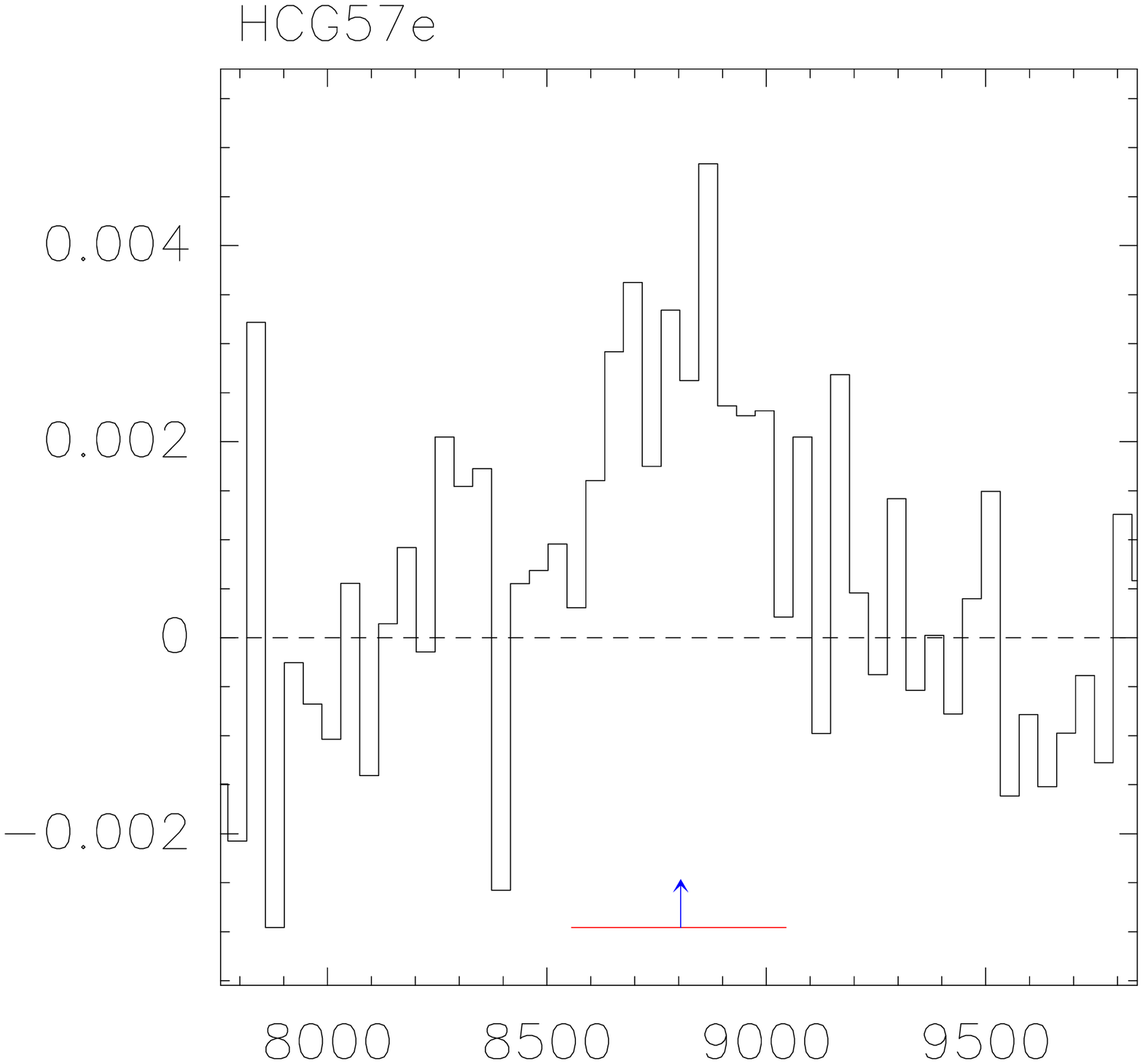}
\hspace{0.1cm}
\includegraphics[width=3.6cm]{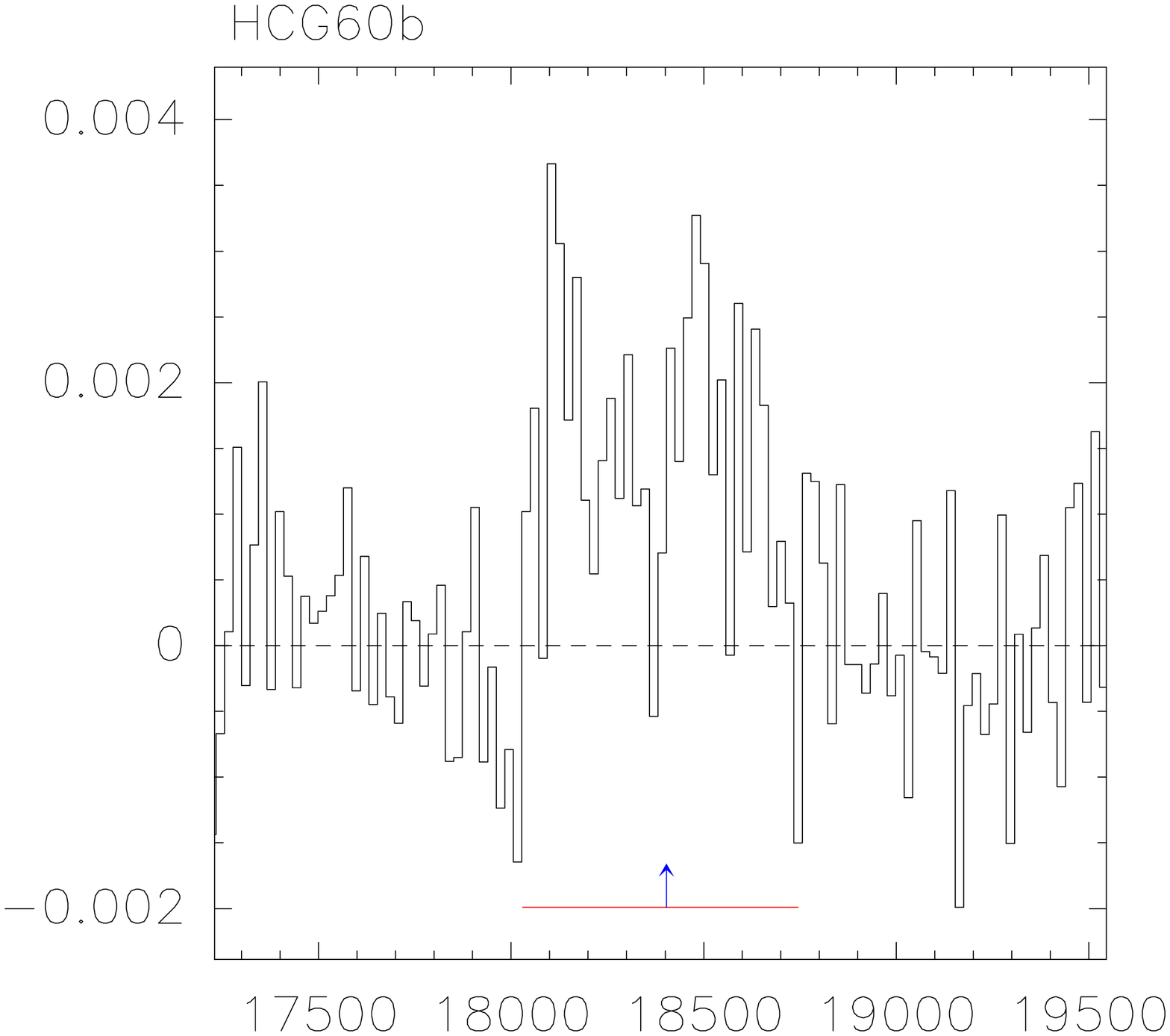}
\hspace{0.1cm}
\includegraphics[width=3.6cm]{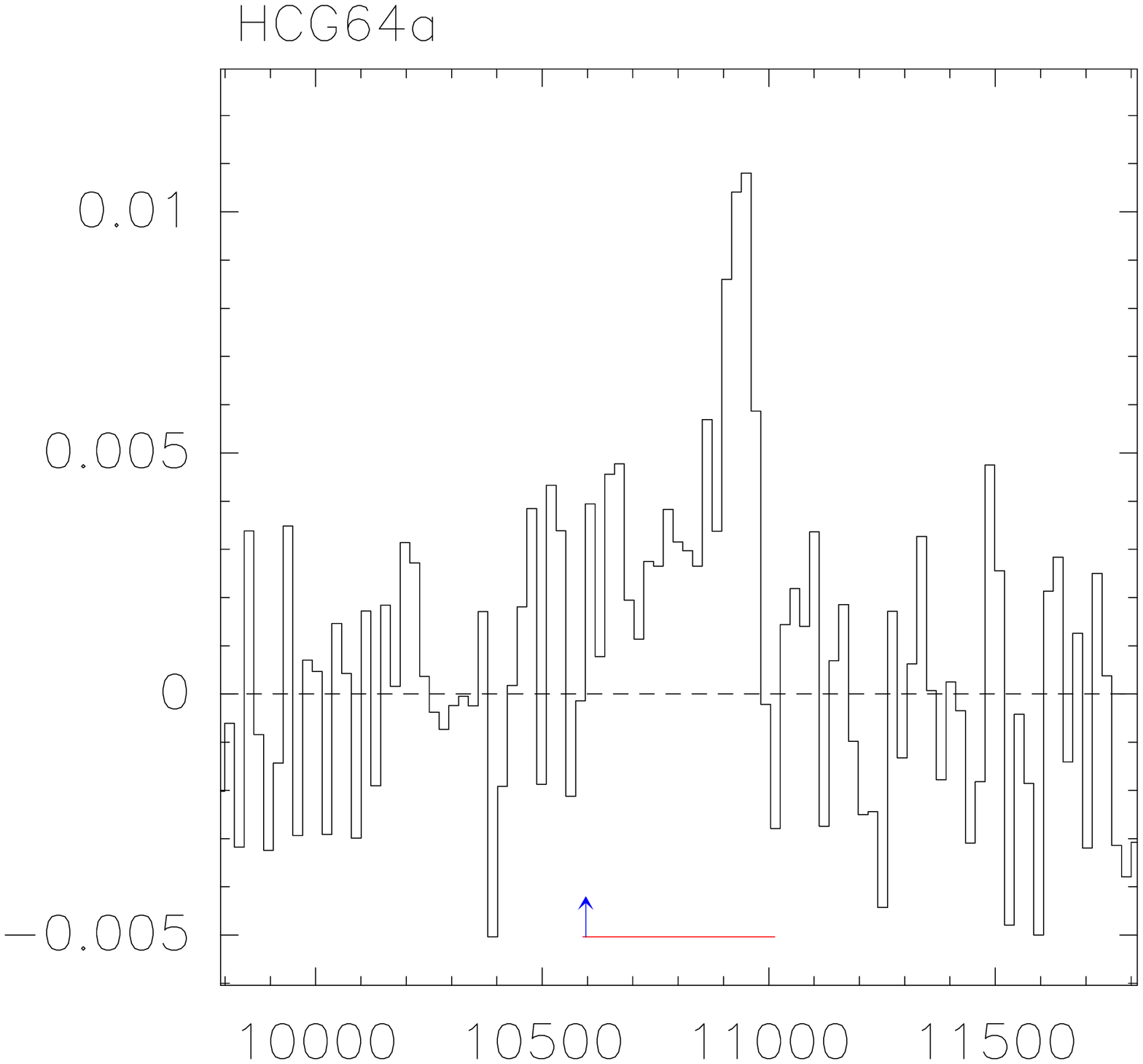}
\includegraphics[width=3.6cm]{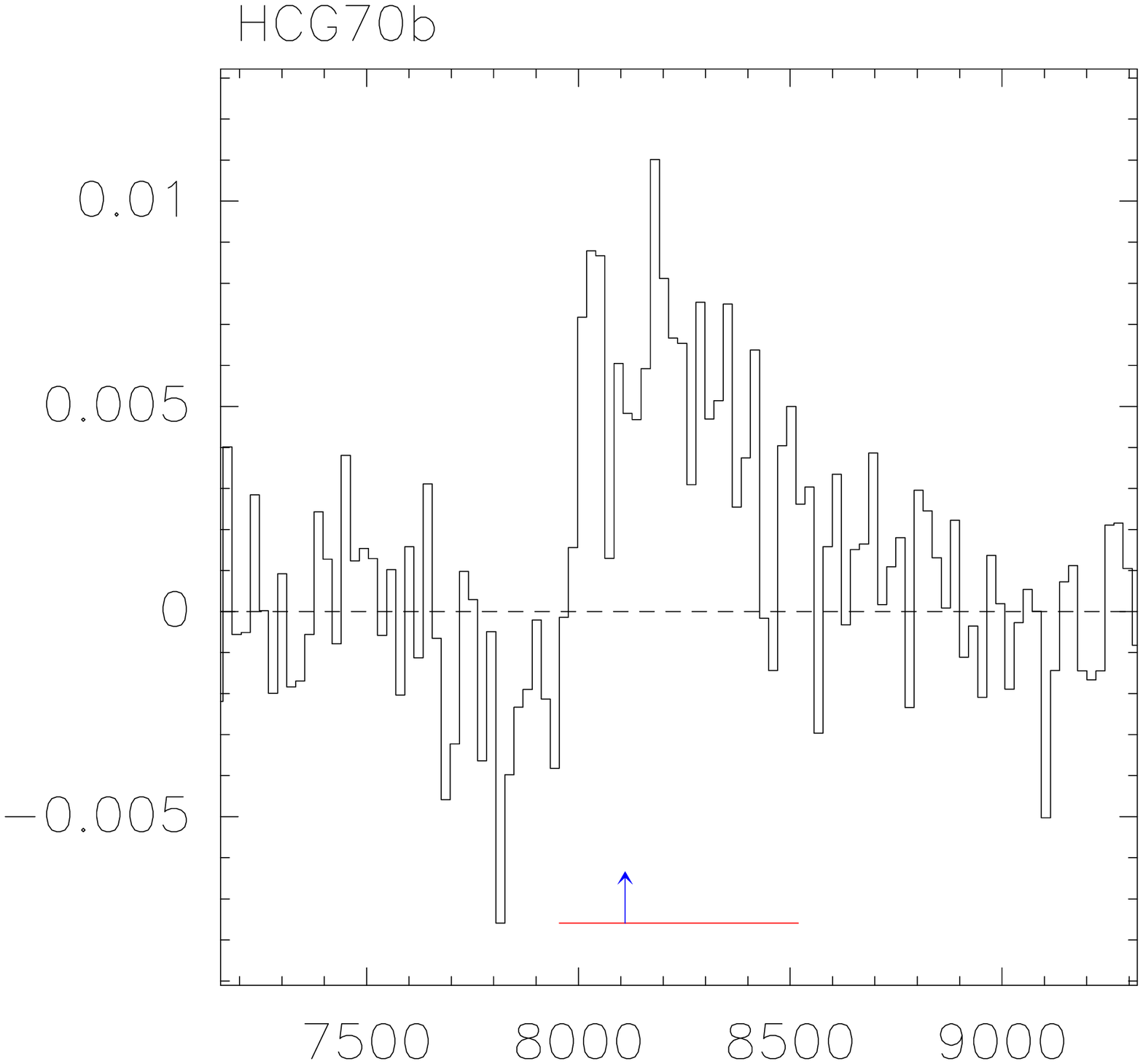}
\hspace{0.1cm}
\hspace{0.1cm}
}
\vspace{0.3cm}
\centerline{
\includegraphics[width=3.6cm]{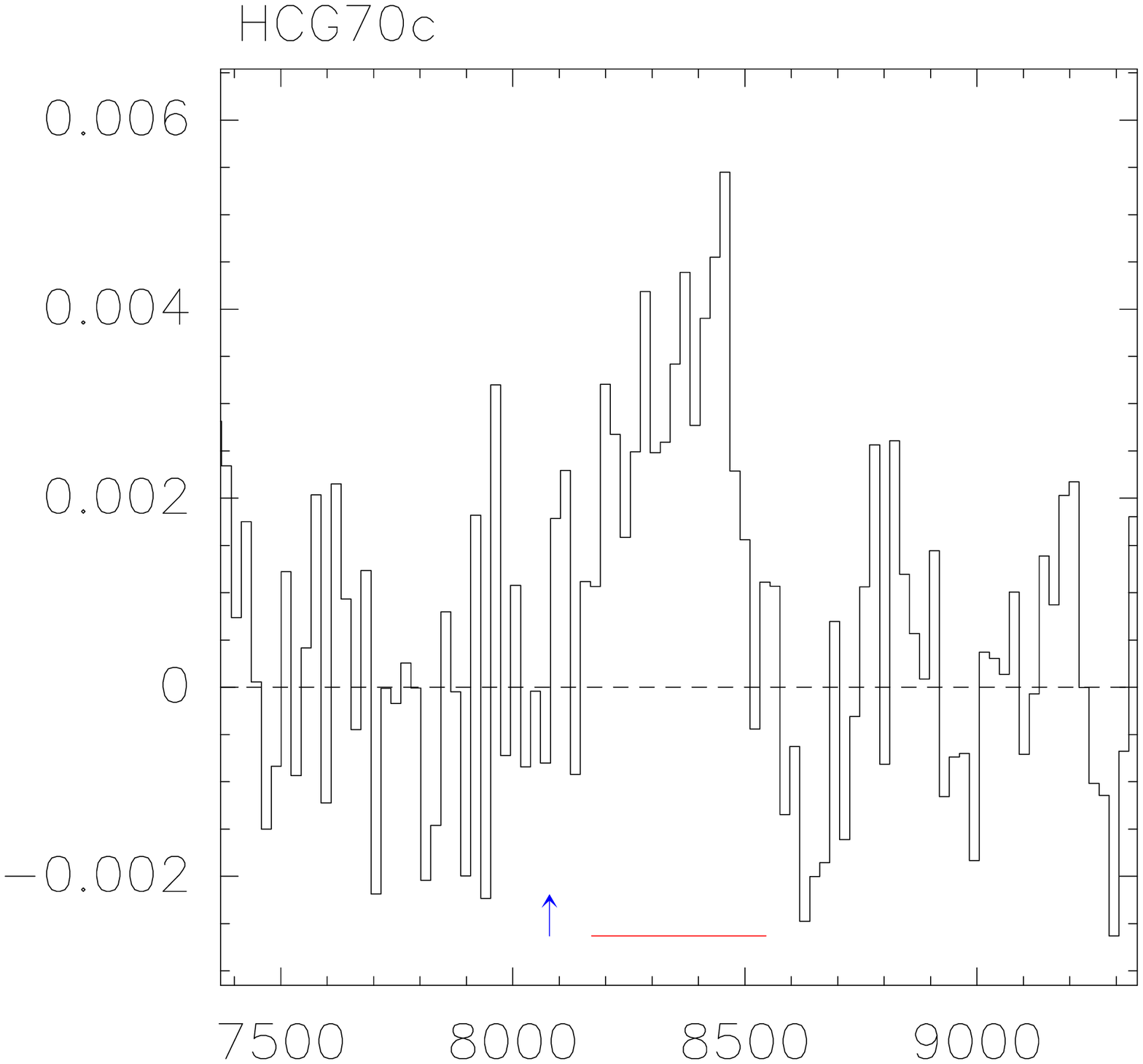}
\hspace{0.1cm}
\includegraphics[width=3.6cm]{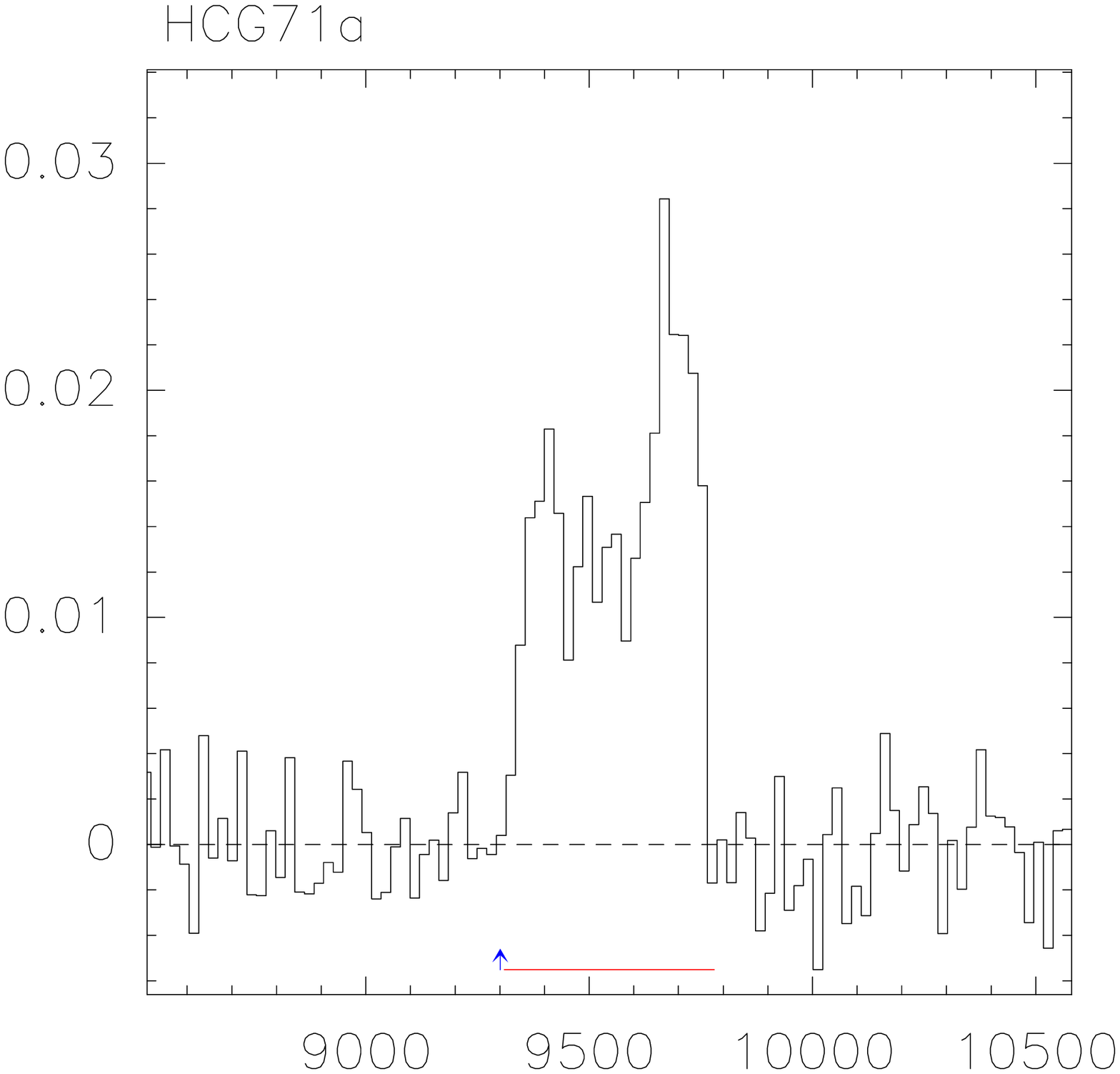}
\hspace{0.1cm}
\includegraphics[width=3.6cm]{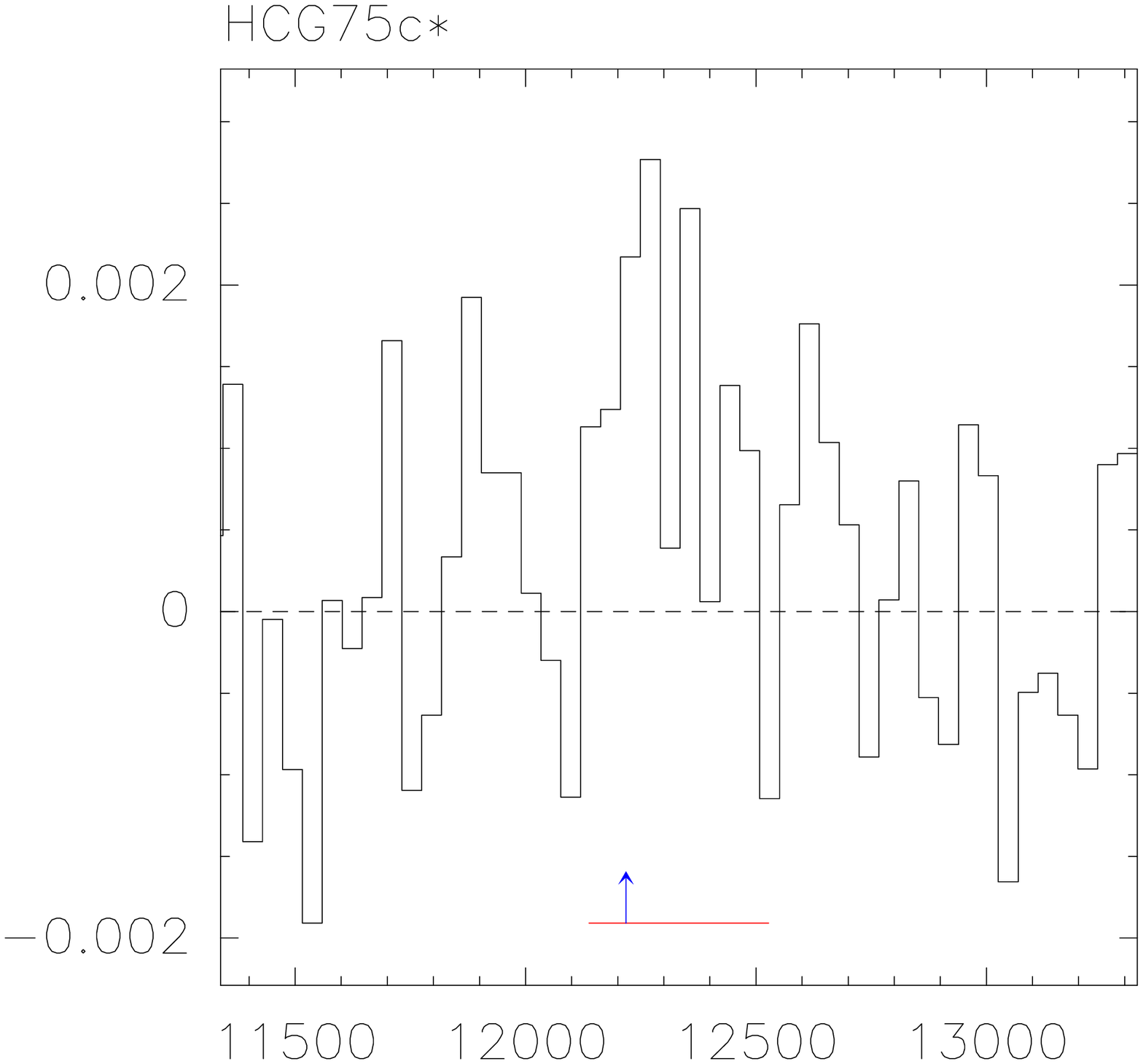}
\hspace{0.1cm}
\includegraphics[width=3.6cm]{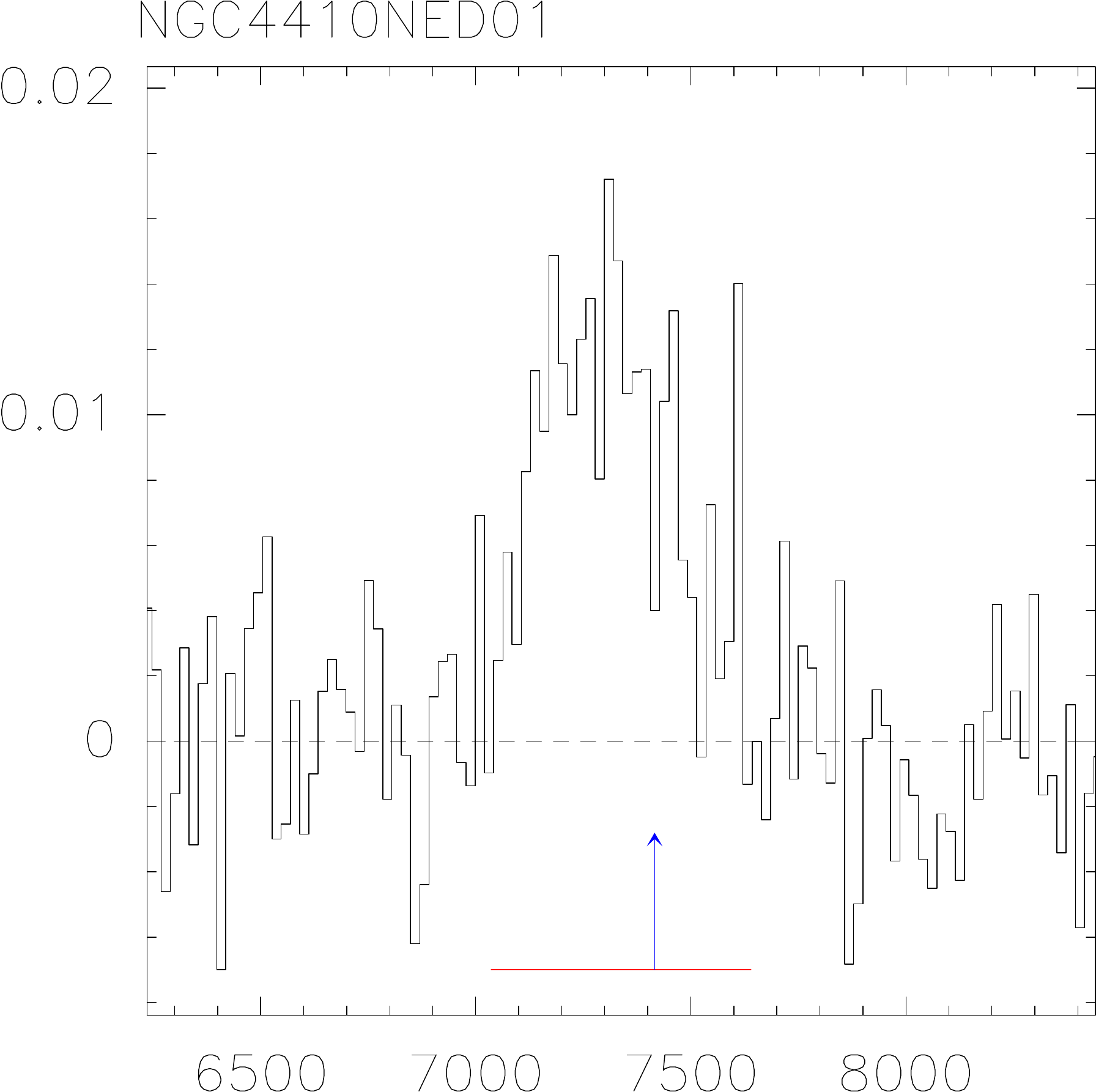}
}
\vspace{0.3cm}
\hspace{1.6cm}
\includegraphics[width=4cm]{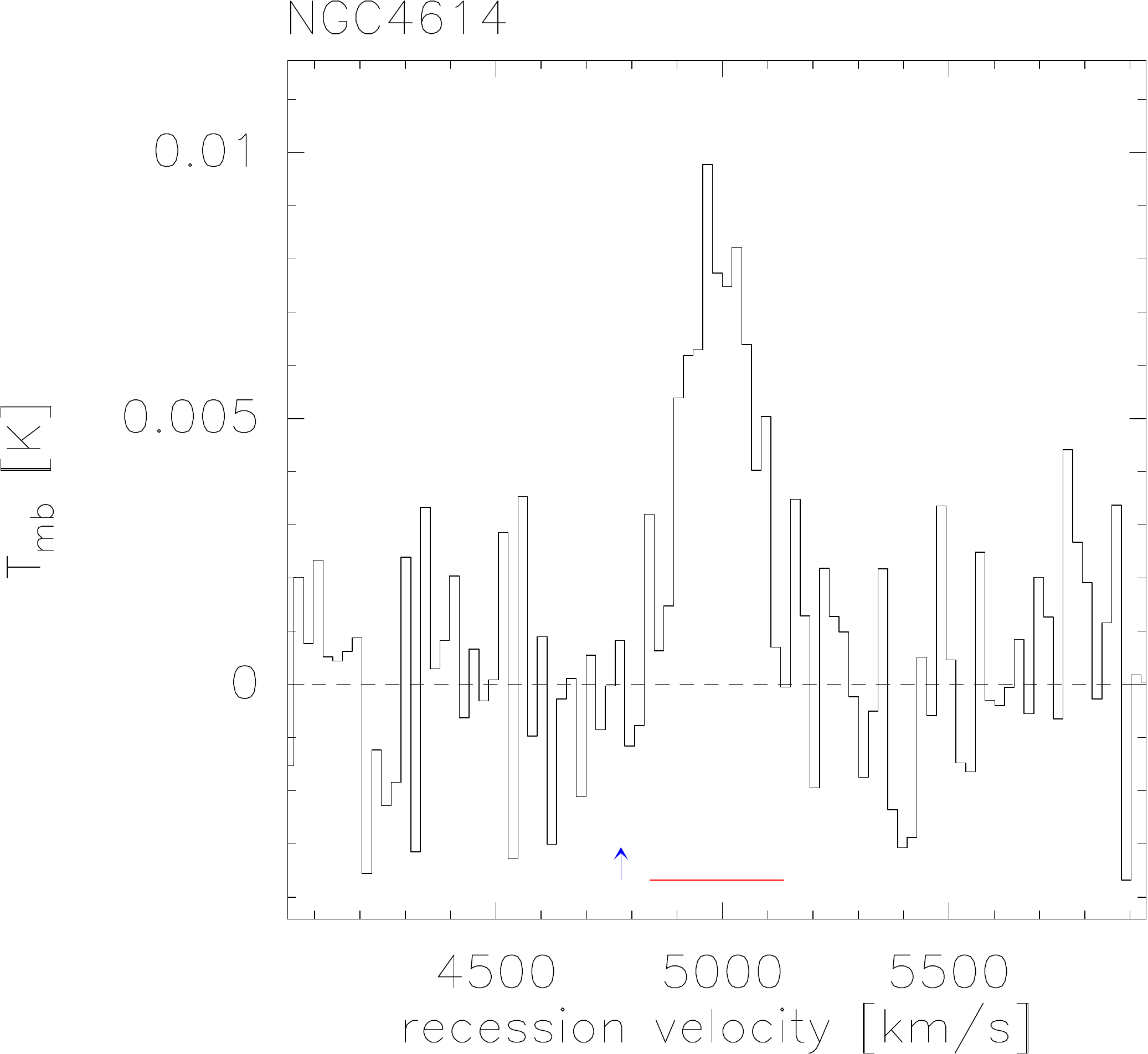}
\hspace{0.1cm}
\includegraphics[width=3.6cm]{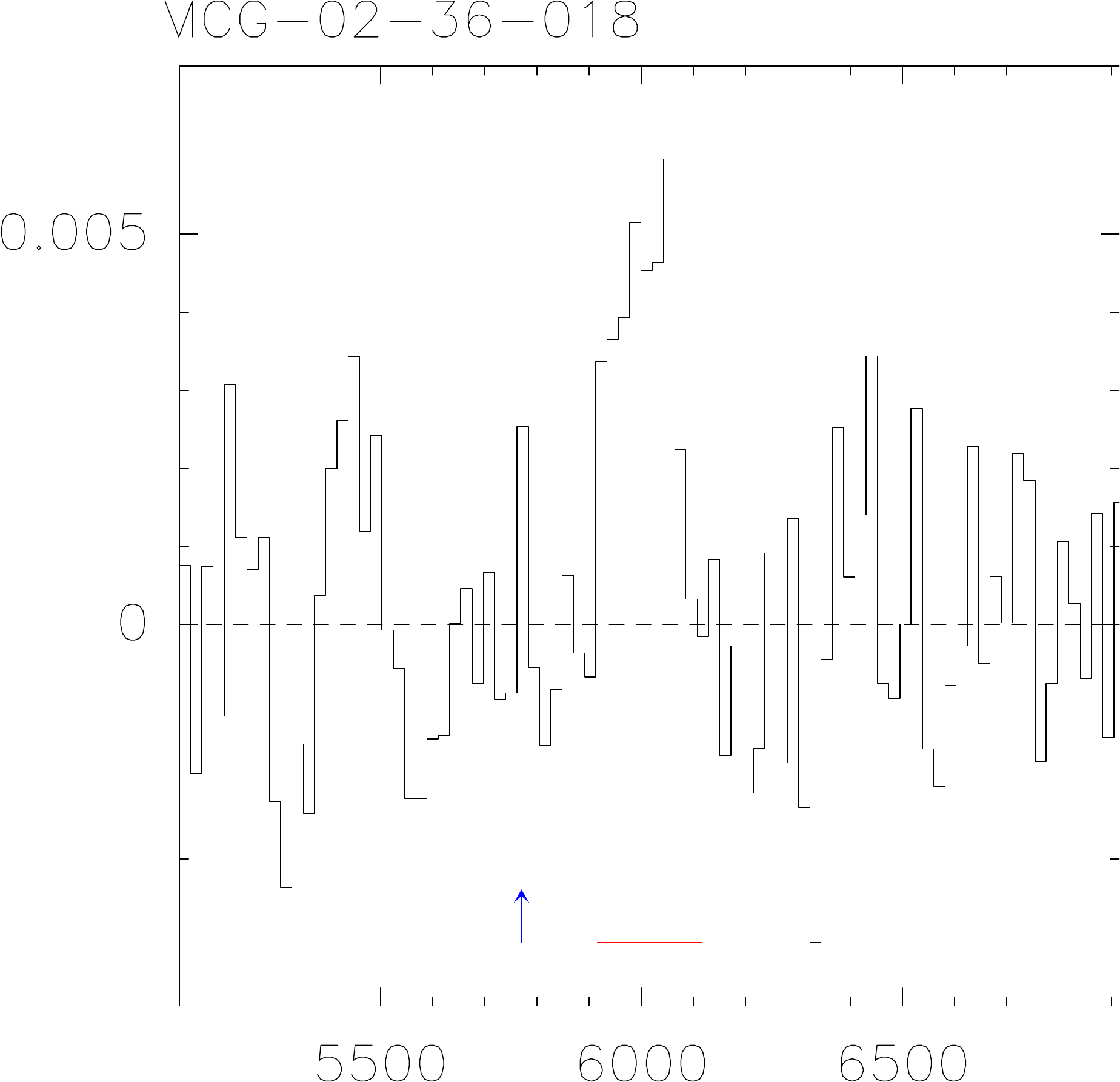}
\caption{CO(1-0) spectra of the detected  spectra (including tentative detections). The velocity resolution
is $\sim$ 20 \kms\ for most spectra and $\sim$  40 \kms\ for some cases where a lower resolution was required
to clearly see the line. The red line segment shows the zero-level linewidth of the 
CO line adopted for the determination of the velocity integrated intensity. 
The blue upright arrow indicated the optical  heliocentric recession velocity. 
An asterisk next to the name indicates a tentative detection.
}
\label{fig:spectra_co10}
\end{figure*}

{ 
\section{Results of a WISE-selected subsample}

As outlined in Sect.~\ref{sec:sample_data}, we
carried out the entire analysis for a WISE  selected subsample to check whether the CO-driven selection has produced a bias.
We selected this subsample by including only those groups in which
all galaxies with reliable WISE data (here S/N $> 2.5~\sigma$) also had CO data.
This subsample consists of 89 galaxies out of the full 130 galaxy sample used for analysis
in the main body of the text.

In Figs.~\ref{fig:mmol_over_mstar_vs_morf_sub_wise} and \ref{fig:sfe_vs_morf_sub_wise} we show the SFE and molecular gas content as a function
of morphological type  for this subsample, analogous to Figs.~\ref{fig:mmol_over_mstar_vs_morf} and \ref{fig:sfe_vs_morf} for the
full sample. The trend seen for the full sample is also present for the WISE-selected
subsample. In Tab.~\ref{tab:mean_mmol_over_mstar_sub_wise} and \ref{tab:mean_sfe_sub_wise} 
we list the results for the mean and median values for
the different subgroups, which entirely confirm the results for the full sample.

   \begin{figure}
   \centering
\includegraphics[width=8cm]{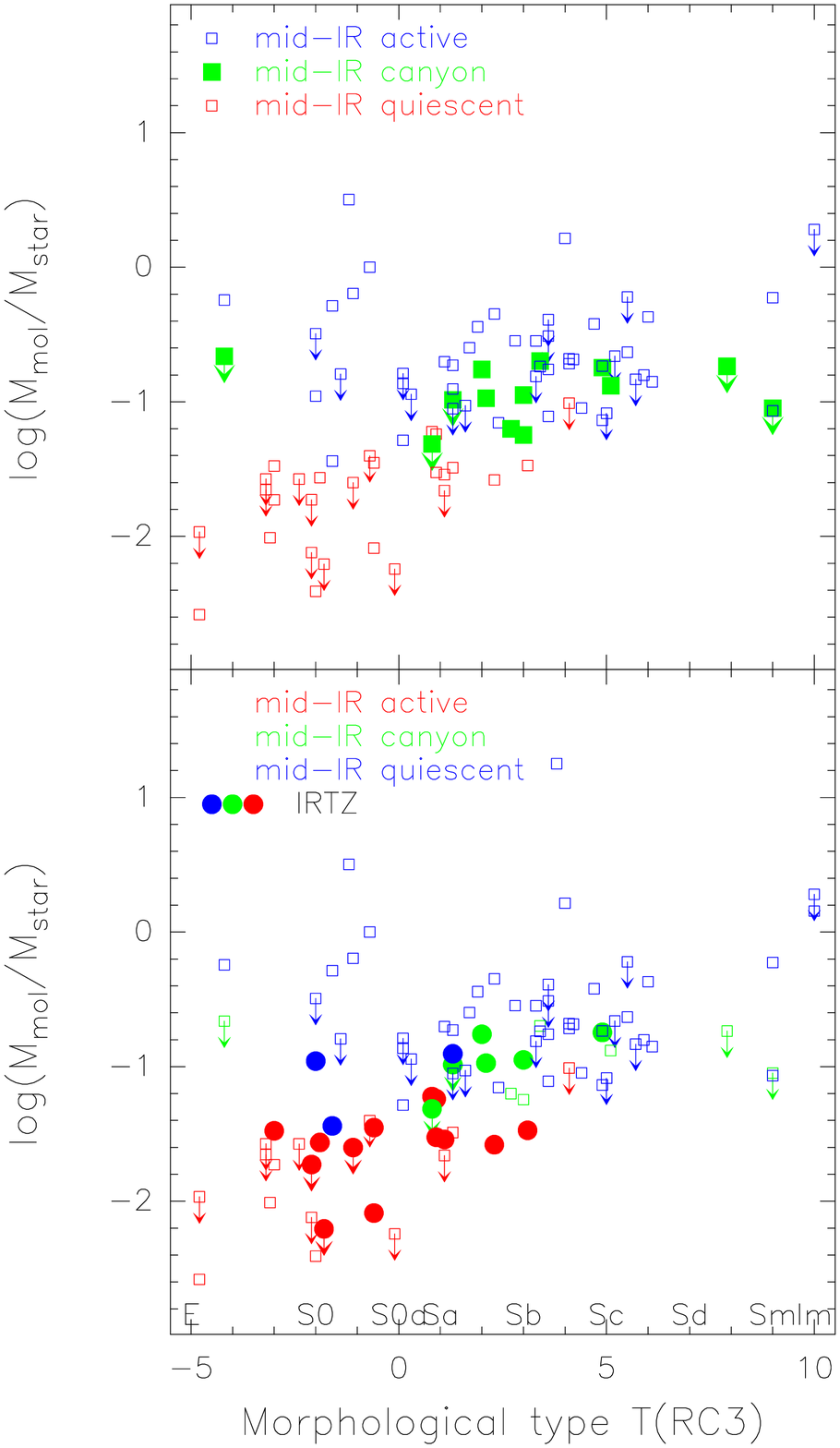}
      \caption{Ratio of molecular gas mass to stellar mass as a function of morphological type
      for the WISE-selected subsample of 89 galaxies with both high-quality WISE and CO data.
       The color coding is as in Fig.~1. 
              Filled circles (lower panel) denote galaxies belonging to the IRTZ . 
              }
   \label{fig:mmol_over_mstar_vs_morf_sub_wise}
   \end{figure}

   \begin{figure}
   \centering
 \includegraphics[width=8cm, clip]{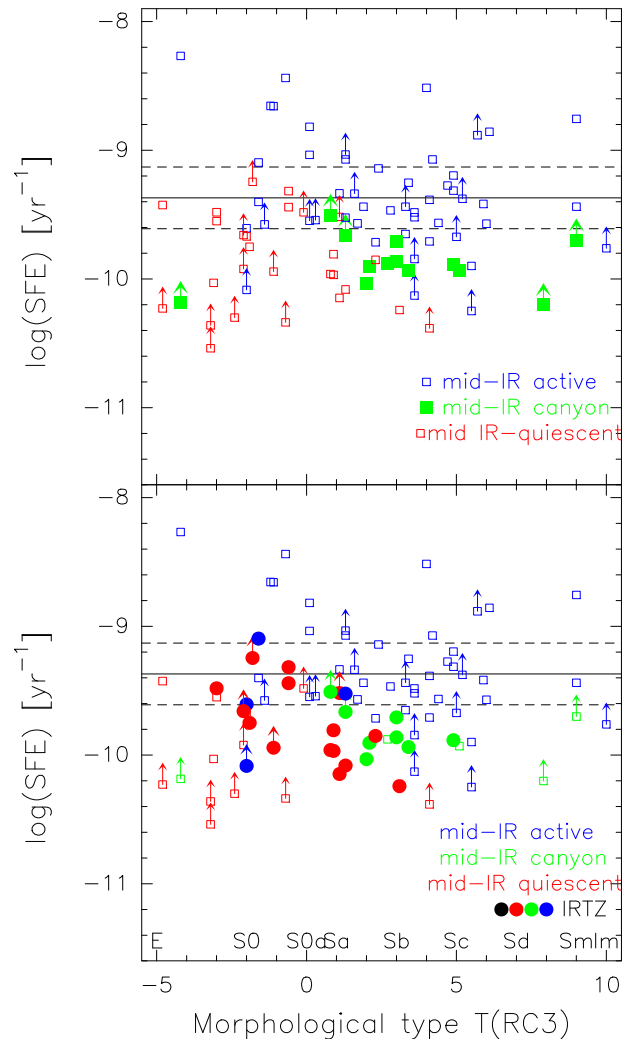}
      \caption{SFE  as a function of morphological type for the WISE-selected subsample
      of 89 galaxies with both high-quality WISE and CO data.
        The color coding is as in Fig.~1. 
     Filled circles denote galaxies belonging to the IRTZ (lower panel). 
      The lines show the
      mean value (full line) and dispersion (dashed lines) found by \citet{bigiel11} for a sample of spiral and starburst galaxies. 
                  }
         \label{fig:sfe_vs_morf_sub_wise}
   \end{figure}

\begin{table}
\caption{Mean and median log(\mmol/\mstar) for different groups of the WISE-selected subsample}
\begin{tabular}{lccc}
\hline
Class & mean & median & n/n$_{\rm up}$\tablefootmark{a} \\
\hline
\multicolumn{3}{l}{WISE-selected subsample, all morphological types}  \\
Active    & -0.76 $\pm$ 0.06  & - 0.77 &49/14\\
Canyon  &  -1.03$\pm$ 0.07 & -1.10& 13/5\\
IRTZ & -1.44$\pm$ 0.10 & -1.48 & 22/5\\
quiescent  &  -1.96 $\pm$ 0.10 & -2.05 & 27/12  \\
\hline
\multicolumn{3}{l}{WISE-selected subsample,  late types  ($T>0$)}  \\
Active    & -0.76 $\pm$ 0.06  & -0.75& 36/10\\
Canyon  &  -1.03$\pm$ 0.07 & -1.10 & 12/4 \\
IRTZ & -1.22$\pm$ 0.09 & -1.31& 13/2 \\
quiescent  &  -1.47 $\pm$ 0.05 & -1.52& 9/2 \\
\hline 
\end{tabular}
\tablefoot{
\tablefoottext{a} {Total number of galaxies (n) and number of upper limits (n$_{\rm up}$)}.
}
\label{tab:mean_mmol_over_mstar_sub_wise}
\end{table}

\begin{table}
\caption{Mean and median log(SFE) (in units of yr$^{-1}$)  for different groups of the WISE-selected subsample}
\begin{tabular}{lccc}
\hline
Class & mean & median & n/n$_{\rm up}$\tablefootmark{a} \\
\hline
\multicolumn{3}{l}{WISE-selected subsample, all morphological types}  \\
Active    & -9.17 $\pm$ 0.07  & - 9.28 &49/4 \\
Canyon  &  -9.79$\pm$ 0.06 & -9.89& 13/5\\
IRTZ & -9.66$\pm$ 0.08 & -9.77 & 22/5\\
quiescent  &  -9.67$\pm$ 0.07 & -9.68 & 27/12  \\
\hline
\multicolumn{3}{l}{WISE-selected subsample,  late types  ($T>0$)}  \\
Active    & -9.28 $\pm$ 0.07  & -9.36& 36/10\\
Canyon  &  -9.79$\pm$ 0.06 & -9.89 & 12/4 \\
IRTZ & -9.86$\pm$ 0.06 & -9.90& 13/2 \\
quiescent  &  -9.95 $\pm$ 0.08 k& -9.97& 9/2 \\
\hline 
\end{tabular}
\tablefoot{
\tablefoottext{a} {Total number of galaxies (n) and number of upper limits (n$_{\rm up}$)}.
}
\label{tab:mean_sfe_sub_wise}
\end{table}


\section{Effect of the aperture correction}
\label{sect:aperture_corr}

We applied an aperture correction, \faper , to our CO measurements
that is  based on the ratio between optical diameter and beam size.
In order to  test whether this aperture correction  could
 have produced a bias in the results, we carried out two tests. 
 
 We  compared the mean and median value of \faper\  for the different galaxy groups to
 search for possible systematic differences that might cause a bias.
 Table~\ref{tab:mean_faper} lists the results and shows that the mean and median values are 
 comparable for the four groups.
 The histogram of the distributions, shown in Fig.~\ref{fig:histo_faper},
 confirms this result and shows that all groups have a similar distribution
 that is strongly skewed toward values of \faper$<2$. There are
 differences, most noticeably between active and canyon galaxies, with
 the former having a  maximum for galaxies $1.25<$ \faper $<1.5$, whereas
 canyon galaxies have more galaxies in the  bin $1.5<$ \faper $<1.75$.
This difference is however too small to produces a serious  bias in our
results. 

As a second test, we carried out our analysis for a subsample with a small
aperture correction, \faper  $< 1.6$, but with a large
enough number of galaxies to obtain statistically significant results.
The distribution of  \mmol/\mstar\  and the SFE as a function of morphological type 
are shown in Figs.~\ref{fig:mmol_over_mstar_vs_morf_faper}
and \ref{fig:sfe_vs_morf_faper}, and the results for the corresponding mean and
median values are shown in Tables \ref{tab:mean_mmol_over_mstar_faper}  
and \ref{tab:mean_sfe_faper}.
The results are entirely consistent with those of the total sample, showing
that the application of the aperture correction has not introduced any biases.

   \begin{figure}
   \centering
 \includegraphics[width=8cm]{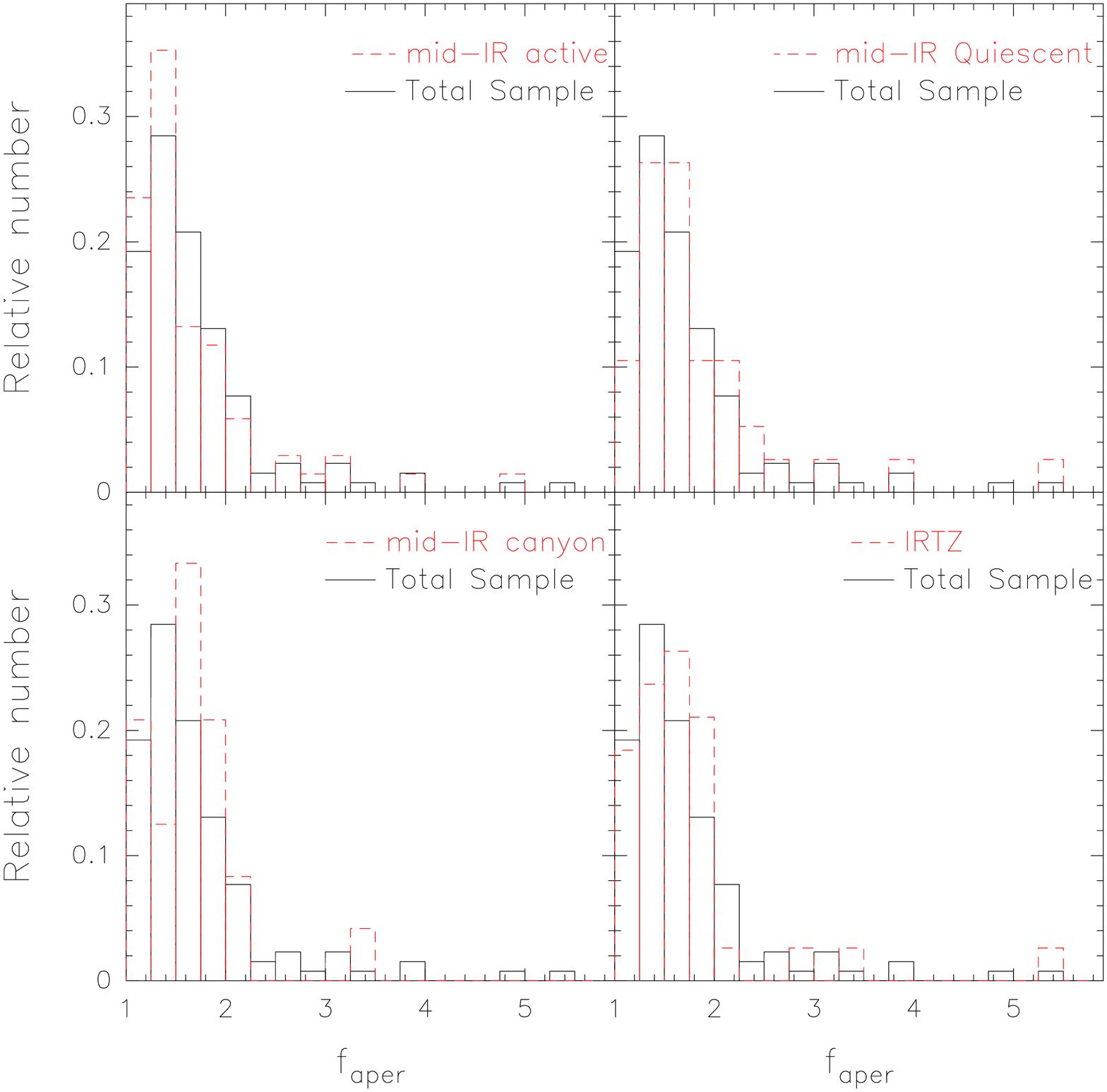}
      \caption{Histogram of the distribution of the aperture correction, \faper , for the
      different subsamples (red dashed line) compared to the full sample (full black line).      }
         \label{fig:histo_faper}
   \end{figure}

\begin{table}
\caption{Mean and median value of the aperture correction, \faper}
\begin{tabular}{lccc}
\hline
Class & mean & median & n\tablefootmark{a} \\
\hline
Total  & 1.70 $\pm$ 0.06 & 1.51 &  130 \\
Active    &  1.64$\pm$0.08  & 1.40 &68 \\
Canyon  &  1.65$\pm$0.10 & 1.59 & 24\\
IRTZ & 1.73$\pm$0.12  & 1.58 & 38\\
Quiescent  &  1.85 $\pm$ 0.13 & 1.60 & 38  \\
\hline 
\end{tabular}
\tablefoot{
\tablefoottext{a} {Total number of galaxies (n)}
}
\label{tab:mean_faper}
\end{table}

   \begin{figure}
   \centering
\includegraphics[width=8cm]{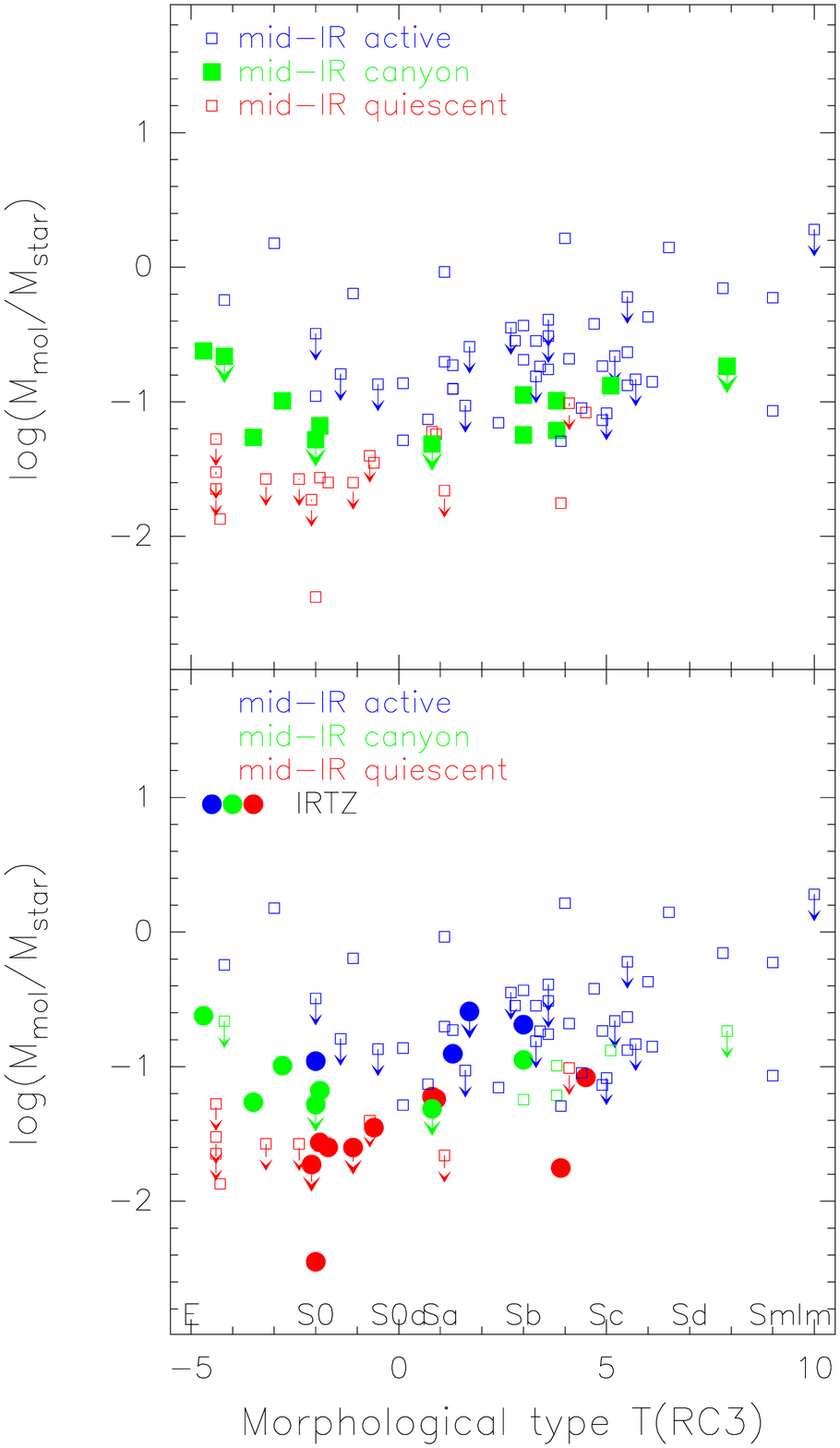}
      \caption{Ratio of molecular gas mass to stellar mass as a function of morphological type
      for galaxies with \faper $<1.6$.
      The color coding is as in Fig.~1. 
        Filled circles (lower panel) denote galaxies belonging to the IRTZ . 
              }
   \label{fig:mmol_over_mstar_vs_morf_faper}
   \end{figure}

   \begin{figure}
   \centering
 \includegraphics[width=8cm]{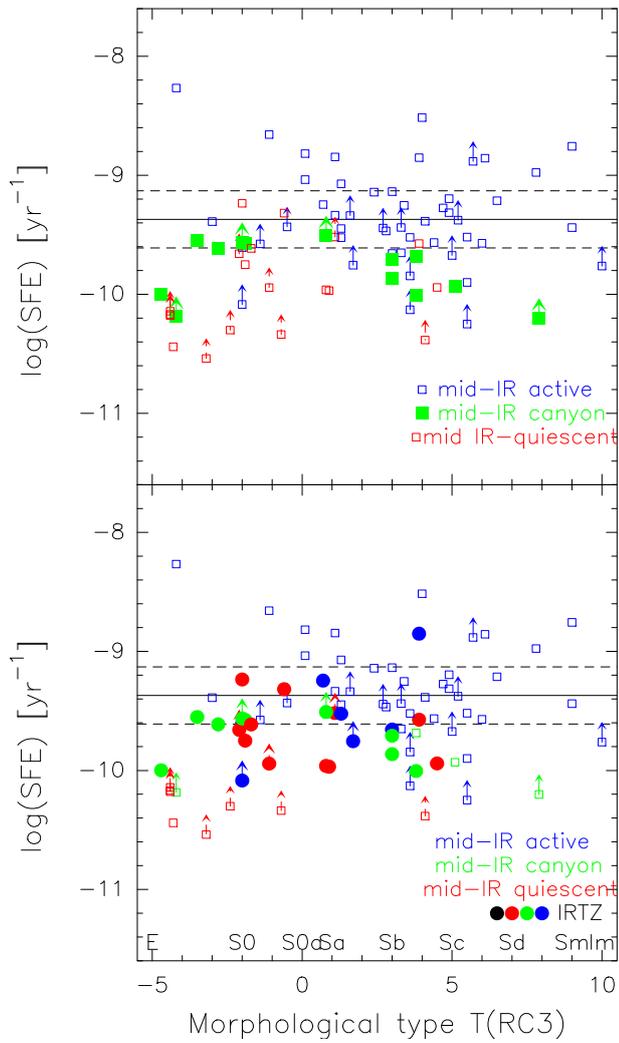}
      \caption{SFE  as a function of morphological type for galaxies with \faper $<1.6$.
      The objects are coded in color according to the classification of \citet{zucker16}.
      The color coding is as in Fig.~1. 
       The lines show the
      mean value (full line) and dispersion (dashed lines) found by \citet{bigiel11} for a sample of spiral and starburst galaxies. 
                  }
         \label{fig:sfe_vs_morf_faper}
   \end{figure}

\begin{table}
\caption{Mean and median log(\mmol/\mstar) for  galaxies with \faper $< 1.6$}
\begin{tabular}{lccc}
\hline
Class & mean & median & n/n$_{\rm up}$\tablefootmark{a} \\
\hline
\multicolumn{3}{l}{Galaxies with \faper $< 1.6$, all morphological types}  \\
Active    & -0.74 $\pm$ 0.06  & - 0.85 &48/14 \\
Canyon  &  -1.20$\pm$ 0.06 & -1.20& 13/4\\
IRTZ & -1.39$\pm$ 0.13 & -1.26 & 21/5\\
quiescent  &  -1.78 $\pm$ 0.14 & -1.82 & 19/10  \\
\hline
\multicolumn{3}{l}{Galaxies with \faper $< 1.6$,  late types  ($T > 0 $)}  \\
Active    & -0.74 $\pm$ 0.06  & -0.76& 39/11\\
Canyon  &  -1.10$\pm$ 0.07 & -1.21 & 7/2 \\
IRTZ & -1.20$\pm$ 0.13 & -1.22& 9/2 \\
quiescent  &  -1.41 $\pm$ 0.13 & -1.50& 6/2 \\
\hline 
\end{tabular}
\tablefoot{
\tablefoottext{a} {Total number of galaxies (n) and number of upper limits (n$_{\rm up}$)}.
}
\label{tab:mean_mmol_over_mstar_faper}
\end{table}

\begin{table}
\caption{Mean and median log(SFE) (in units of yr$^{-1}$) for galaxies with \faper $< 1.6$}
\begin{tabular}{lccc}
\hline
Class & mean & median & n/n$_{\rm up}$ \tablefootmark{a}\\
\hline
\multicolumn{3}{l}{Galaxies with \faper $< 1.6$, all morphological types}  \\
Active    & -9.17$\pm$ 0.06  & - 9.25 &48/14 \\
Canyon  &  -9.72$\pm$ 0.06 & -9.70& 13/4\\
IRTZ & -9.61$\pm$ 0.05 & -9.62 & 21/5\\
quiescent  &  -9.63$\pm$ 0.09 & -9.64 & 19/10  \\
\hline
\multicolumn{3}{l}{Galaxies with \faper $< 1.6$,  late types  ($T > 0$)}  \\
Active    & -9.23 $\pm$ 0.06 & -9.26& 39 /11\\
Canyon  &  -9.78$\pm$ 0.07 & -9.86 & 7/2 \\
IRTZ & -9.73$\pm$ 0.06 & -9.78& 9/2 \\
quiescent  &  -9.79 $\pm$ 0.09 & -9.95& 6/2 \\
\hline 
\end{tabular}
\tablefoot{
\tablefoottext{a} {Total number of galaxies (n) and number of upper limits (n$_{\rm up}$)}.
}
\label{tab:mean_sfe_faper}
\end{table}

}
\end{document}